# Distributive Network Utility Maximization (NUM) over Time-Varying Fading Channels

Junting Chen, Vincent K. N. Lau, *Senior Member, IEEE,* and Yong Cheng, *Student Member, IEEE*

***Abstract*—Distributed network utility maximization (NUM) has received an increasing intensity of interest over the past few years. Distributed solutions (e.g., the primal-dual gradient method) have been intensively investigated under fading channels. As such distributed solutions involve iterative updating and explicit message passing, it is unrealistic to assume that the wireless channel remains unchanged during the iterations. Unfortunately, the behavior of those distributed solutions under *time-varying* channels is in general unknown. In this paper, we shall investigate the convergence behavior and tracking errors of the iterative primal-dual scaled gradient algorithm (PDSGA) with dynamic scaling matrices (DSC) for solving distributive NUM problems under time-varying fading channels. We shall also study a specific application example, namely the multi-commodity flow control and multi-carrier power allocation problem in multi-hop ad hoc networks. Our analysis shows that the PDSGA converges to a limit region rather than a single point under the finite state Markov chain (FSMC) fading channels. We also show that the order of growth of the tracking errors is given by $\mathcal{O}\left(\overline{T}/\overline{N}\right)$, where $\overline{T}$ and $\overline{N}$ are the update interval and the average sojourn time of the FSMC, respectively. Based on this analysis, we derive a low complexity distributive adaptation algorithm for determining the adaptive scaling matrices, which can be implemented distributively at each transmitter. The numerical results show the superior performance of the proposed dynamic scaling matrix algorithm over several baseline schemes, such as the regular primal-dual gradient algorithm.**

## I. Introduction

The Network Utility Maximization (NUM) framework has been widely adopted for network resource allocation problems in wireline and wireless networks over the past few years. In such NUM problems, the network control problem is formulated as an optimization problem, in which the utility function of the network (e.g., efficiency, user satisfactory, etc.) is to be maximized, under some resource constraints (e.g., power constraint, link capacity constraint, etc.) [1]–[7]. There are a lot of works that focus on solving the NUM problems (e.g., see [4] for a comprehensive survey). Based on the primal-dual decomposition framework [2], [3], distributive algorithms such as the dual-gradient algorithm [1]–[7] and the primal-dual gradient algorithm [8]–[10], have been proposed. These distributive algorithms are quite useful for practical networks, which consist of randomly placed nodes and lack of centralized coordination. However, the implementation of the distributive algorithms involves *iterative solutions with explicit message passing* among nodes, which means that the algorithm may not converge within a negligible time especially in the wireless networks. In the existing literature [1]–[7], the convergence and the optimality properties of these distributive algorithms are established under the *quasi-static* fading channel assumption, i.e., the channel coefficients between any pair of nodes in the network are assumed to be *constant* during the updates of the iterative algorithms. However, since the wireless channel is time-varying by nature and explicit message passing between network nodes is required for each iteration, it is a limiting assumption to assume that the channels remain unchanged during a significant number of iterative steps before the algorithm converges [11]. As a result, it is of great importance both theoretically and practically to study the behavior of the distributive algorithms under time-varying channels. In this paper, we are interested in the behavior of the generalized *primal-dual scaled gradient algorithm* [8]–[10] for solving the distributed NUM problem in multihop wireless networks under time-varying channels (e.g., finite-state Markov channel). For such multihop wireless networks with time-varying channels, the optimal solution of the NUM problem, which depends on the channel state information (CSI) across the network, would also be time-varying. Hence, there are some technical challenges which require further investigation in order to understand the behavior of these distributive iterative algorithms in time-varying CSI situations. For example,

- **How to quantify the performance penalty due to time-varying channels?** Most of the NUM algorithms are designed assuming the channel is static during the iterations of the algorithms. Hence, it is important to have a better understanding of the performance penalty due to the time-varying channels. For example, due to the time-varying channels, it is not possible to guarantee that the problem constraints (such as the power constraints or flow balance constraints) will be satisfied at every time slot.
- **What is the cost-performance tradeoff?** It is also important to study the tradeoff relationship between how fast the iterative algorithm updates (the message passing overhead), the speed of the time varying channels and the performance penalty of the NUM problem. However, doing so involves careful analysis of the transient of the iterative algorithms, which is highly non-trivial.

The authors are with the Department of Electronic and Computer Engineering (ECE), The Hong Kong University of Science and Technology (HKUST), Hong Kong (e-mail: eejtchen@ust.hk; chengy@ust.hk; eeknlau@ust.hk).

- **How to enhance the existing NUM solutions?** Furthermore, based on the understanding of the performance penalty, we are interested to enhance the existing iterative NUM algorithms (designed for quasi-static channels) to improve the associated tracking performance in time-varying channels.

Due to the stochastic nature of the wireless channel and the nonlinear dynamics of the iteration processes encountered, it is *highly nontrivial* to answer the above questions in general [11]. There are some preliminary works which study the impacts of time-varying inputs to the algorithm. In Kelly's classical work [12], the author introduced stochastic perturbations in the algorithm to represent random load entering the network. In [5], the authors studied a stochastic NUM driven by the stochastic noisy feedback. However, [12] and [5] did not consider the impact of time-varying CSI and their system models have a unique equilibrium point. To study the impacts of time-varying CSI on distributive algorithms, the authors of [11] have studied the performance of the *distributed scaled gradient projection algorithm* for tracking the *moving* Nash Equilibrium (NE) of the multicarrier interference game under the finite-state Markov channel (FSMC) model, based on *randomly switched system modeling* [13]. In [14], the *hybrid system* model was used to study the multicell CDMA interference game. While these works [11], [14] provide some preliminary results on the behavior of the distributed algorithms for solving games under time-varying channels, to the best of our knowledge, there is no related work studying the behavior of distributed algorithms for NUM problems under time-varying CSI. Furthermore, due to the decomposition techniques [2], [3], the dynamics of the distributed algorithms for the NUM problems are quite different to those for solving non-cooperative games. On the other hand, there are also some works on estimation theory concerning parameter tracking in nonstationary environments [15], [16]. However, the techniques and results of [15], [16], which are based on the special linear structure of the underlying dynamics (e.g., linear regression model and the linear least square estimates), cannot be applied to the general distributive NUM problems we considered in this paper.

In this paper, we shall attempt to shed some light on the above technical challenges by studying the behavior of an iterative primal-dual scaled gradient algorithm (PDSGA) [8]–[10] under time-varying CSI. We model the CSI by a finite state Markov chain (FSMC) [17]–[19]. Using sample-path analysis on the *algorithm trajectory*, we derived closed-form expressions for the algorithm tracking errors, which are defined as the difference between the algorithm trajectory and the moving target optimal solution of the associated NUM problem at a given time. Due to the time varying channels, the NUM problem constraints may not be satisfied at every time slot and we define such event as *constraint outage*. We have also quantified the probability of constraint outages in terms of the speed of the time-varying CSI as well as the *constraint backoff margins*. Based on these results, we shall propose a novel algorithm to dynamically adjust the scaling matrix in the PDSGA based on the current CSI. The proposed solution has low complexity and can be implemented distributively at each receiver node utilizing the local CSI only. As an illustration, we consider an application example, namely the joint multi-commodity flow control and multi-carrier power allocation (MCFC-MCPA) problem [1], [4]. We compare the performance of the PDSGA with the proposed dynamic scaling matrices against various baseline references.

The paper is organized as follows. In Section II, we introduce the general NUM formulation, FSMC model, and an example. In Section III, we elaborate the primal-dual scaled gradient algorithm (PDSGA) and its convergence behavior. In Section IV, we shall derive low complexity solution of the adaptive scaling matrices. Section V demonstrates the tracking performance of the PDSGA using the proposed dynamic scaling matrix algorithm and verifies the analytical results by simulations. Finally, we conclude with a summary of the main results in Section VI.

*Notations*: Matrix and vectors are denoted with capitalized and small boldface letters, respectively. $\mathbf{A}^T$ ($\mathbf{a}^T$) denotes the transpose of matrix (vector) $\mathbf{A}$ ($\mathbf{a}$). $[\mathbf{A}]_{i,j}$ denotes the $(i,j)^{th}$ entry of matrix $\mathbf{A}$, and $\mathbf{I}_{N_F}$ denotes the $N_F \times N_F$ identity matrix. $\mathbb{C}$, and $\mathbb{R}_+$ denote the set of complex numbers and non-negative real numbers, respectively. $\mathbb{E}$ denotes the operation of taking expectation; $\bigotimes$ denotes the operation of Cartesian product; $\mathbf{a} \succeq \mathbf{b}$ denotes componentwise comparison. Finally, $\mathbf{1}$ denotes a column vector of all ones.

## II. System Model

In this section, we shall introduce the general network utility maximization (NUM) formulation, the finite state Markov channel (FSMC) model, as well as the application example.

### A. Network Utility Maximization under Time-varying Channels

Consider a multihop wireless network with $K$ nodes, with $\mathcal{K}$ denoting the set of all nodes. We shall first consider the following optimization problem with uncoupled utilities and coupled constraints [1]–[7]. We shall illustrate with an application example in Section II-C how this optimization problem corresponds to general NUM problems.

*Problem 1 (General Network Utility Maximization Problem):* At time-slot $t$, the network-wide resource allocation problem is given by

$$\begin{array}{cl} \underset{\mathbf{x}}{\text{maximize}} & \sum_{k=1}^{K} f_k\left(\mathbf{x}_k; \mathbf{h}(t)\right) \\ \text{subject to} & \mathbf{g}\left(\mathbf{x}; \mathbf{h}(t)\right) \succeq \mathbf{0}, \end{array} \tag{1}$$

where $f_k\left(\mathbf{x}_k; \mathbf{h}(t)\right) : \mathbb{R}_+^{M_k} \to \mathbb{R}$ denotes the concave utility function of the $k^{th}$ node; $\mathbf{x}_k \in \mathbb{R}_+^{M_k}$ is the resource allocation vector at the $k^{th}$ node, with $M_k$ dimensions; the vector $\mathbf{x} = \left[\mathbf{x}_1^T \ \mathbf{x}_2^T \ \cdots \ \mathbf{x}_K^T\right]^T \in \mathbb{R}_+^M$ is the collection of all the resource allocation vectors $\{\mathbf{x}_k\}_{k=1}^K$, with $M = \sum_{k=1}^K M_k$; $\mathbf{g}\left(\mathbf{x}; \mathbf{h}(t)\right) = \left[g_1\left(\mathbf{x}; \mathbf{h}(t)\right) \ \ g_2\left(\mathbf{x}; \mathbf{h}(t)\right) \ \cdots \ g_{N_c}\left(\mathbf{x}; \mathbf{h}(t)\right)\right]^T$ are concave functions (which are related to the flow-balance constraints in the wireless networks), with $N_c$ representing the number of inequality constraints; and $\mathbf{h}(t)$ denotes the collection of the global channel fading coefficients (GCSI) of the network.

Such optimization problem in (1) has been widely studied in the literature for quasi-static CSI $\{\mathbf{h}(t)\}$. For instance, the GCSI $\{\mathbf{h}(t)\}$ are assumed to be constants while solving the NUM problem in [1]–[7]. A common approach to solve *Problem 1* (with constant $\mathbf{h}(t)$) is the *dual decomposition* approach [2], [3], which is based on the *Lagrangian multiplier* (LM) theory [20]. Specifically, the *Lagrangian* of *Problem 1* can be written as:

$$\mathcal{L}(\mathbf{x}, \boldsymbol{\lambda}; \mathbf{h}(t)) = \sum_{k=1}^{K} f_k(\mathbf{x}_k; \mathbf{h}(t)) + \boldsymbol{\lambda}^T \mathbf{g}(\mathbf{x}; \mathbf{h}(t)), \quad (2)$$

where $\boldsymbol{\lambda}$ are the LM associated with the inequality constraints.

Before preceding further, we make the following assumptions.

- (A1) $f_k(\mathbf{x}_k; \mathbf{h}(t))$ is a twice-differentiable strictly concave function in $\mathbf{x}_k \succeq \mathbf{0}$, and $\{g_i(\mathbf{x}; \mathbf{h}(t))\}_{i=1}^{N_c}$ are twice-differentiable concave functions in $\mathbf{x} \succeq \mathbf{0}, \forall k \in \mathcal{K}$.
- (A2) There exits a vector $\mathbf{x}(\mathbf{h}(t)) \succeq \mathbf{0}$, such that $\mathbf{g}(\mathbf{x}; \mathbf{h}(t)) \succ \mathbf{0}$.
- (A3) Primal and dual optimal values of *Problem 1* can be attained [20].

Under *Assumption (A1)*, *Problem 1* admits a unique global optimal solution $\mathbf{x}^*(\mathbf{h}(t))$, and under *Assumption (A2)*, strong duality holds for *Problem 1*, i.e., the duality gap is zero [20]. Moreover, the three assumptions together suggest that the Kuhn-Tucker theorem [8]–[10] applies for *Problem 1*, i.e., a vector $\mathbf{x}^*(\mathbf{h}(t)) \succeq \mathbf{0}$ is an optimal solution of *Problem 1*, *if and only if* there is a vector $\boldsymbol{\lambda}^*(\mathbf{h}(t)) \succeq \mathbf{0}$ such that $(\mathbf{x}^*(\mathbf{h}(t)), \boldsymbol{\lambda}^*(\mathbf{h}(t)))$ is a saddle point of the Lagrangian $\mathcal{L}(\mathbf{x}, \boldsymbol{\lambda}; (\mathbf{h}(t)))$ [8]–[10]. As a result, solving the concave programming *Problem 1* and finding the saddle point of the Lagrangian $\mathcal{L}(\mathbf{x}, \boldsymbol{\lambda}; \mathbf{h}(t))$ are equivalent [8]–[10]. In the literature, primal-dual algorithms are widely used to solve *Problem 1*. We shall discuss this through a specific example in Section II-D.

### B. Finite State Markov Channel Model

In this section, we shall elaborate the statistic model for the time-varying CSI. Let $\mathbf{h}_l(t) \in \widetilde{\mathcal{H}}_l$ denotes the channel fading coefficients between the transmitter-receiver pair on the $l^{th}$ link at time-slot $t$, $\forall l \in \mathcal{E}$. Motivated by the simplicity of the FSMC model [17]–[19] for time-varying channels, we model the channel fading process $\left\{\mathbf{h}_l(t) \in \widetilde{\mathcal{H}}_l\right\}$ as an ergodic finite-state Markov chain (FSMC), with state space $\widetilde{\mathcal{H}}_l$ and cardinality $|\widetilde{\mathcal{H}}_l| = \widetilde{Q}_l$, $\forall l \in \mathcal{E}$. For the FSMC $\{\mathbf{h}_l(t)\}$, we make the following assumptions.

- (A4) Similar to [17]–[19], the transition probability matrix $\mathbf{T}_l \in \mathbb{R}_+^{\widetilde{Q}_l \times \widetilde{Q}_l}$ of the FSMC $\{\mathbf{h}_l(t)\}$ is assumed to have the following structure:

$$\mathbf{T}_l = \begin{bmatrix} \nu & \varepsilon & 0 & 0 & 0 & \cdots & 0 & \varepsilon \\ \varepsilon & \nu & \varepsilon & 0 & 0 & \cdots & 0 & 0 \\ 0 & \varepsilon & \nu & \varepsilon & 0 & \cdots & 0 & 0 \\ \vdots & \vdots & \vdots & \vdots & \vdots & \ddots & \vdots & 0 \\ \varepsilon & 0 & 0 & 0 & 0 & \cdots & \varepsilon & \nu \end{bmatrix}, \forall l \in \mathcal{E}, \quad (3)$$

where $\nu = 1 - 2\varepsilon$ and $\varepsilon = \mathcal{O}(f_d \tau)$, with $f_d$ and $\tau$ denoting the doppler frequency shift and symbol duration, respectively.

Let $\mathbf{h}(t) \in \mathcal{H}$ denote the collection of all the channel fading coefficients (GCSI), then $\{\mathbf{h}(t)\}$ is also an ergodic finite-state Markov chain. The *state space* $\mathcal{H}$ and *transition probability matrix* $\mathbf{T}$ of the FSMC $\{\mathbf{h}(t)\}$ are given by

$$\mathcal{H} = \bigotimes_{\{l \in \mathcal{E}\}} \widetilde{\mathcal{H}}_l, \text{ and } \mathbf{T} = \bigotimes_{\{l \in \mathcal{E}\}} \mathbf{T}_l \in \mathbb{R}_+^{Q \times Q}, \quad (4)$$

where $Q = |\mathcal{H}| = \prod_{\{l \in \mathcal{E}\}} \widetilde{Q}_l$ is the cardinality of the aggregate state space $\mathcal{H} = \left\{\mathbf{h}^{(1)}, \mathbf{h}^{(2)}, \cdots, \mathbf{h}^{(Q)}\right\}$ of the FMSC $\{\mathbf{h}(t)\}$. For notational convenience, we shall use $q \in \mathcal{Q} = \{1, 2, \cdots, Q\}$ as the indexing variable for enumerating the realization of $\{\mathbf{h}(t)\}$ in the rest of the paper.

### C. Joint Multicommodity Flow Control and Multicarrier Power Allocation

As an application example, we consider a *joint multicommodity flow control and multicarrier power allocation* (MCFC-MCPA) problem [1], [4] in *Problem 1*. Consider a multihop wireless network modeled by a *directed* graph [1], [4] $\mathcal{G} = (\mathcal{K}, \mathcal{E})$, where $\mathcal{K}$ is the set of nodes and $\mathcal{E}$ is the set of directed edges (i.e. delivering flows for the corresponding commodities), with $|\mathcal{K}| = K$ and $|\mathcal{E}| = L$. Fig. 1 illustrates an example wireless network with $K = 6$ nodes and $L = 8$ directed links.

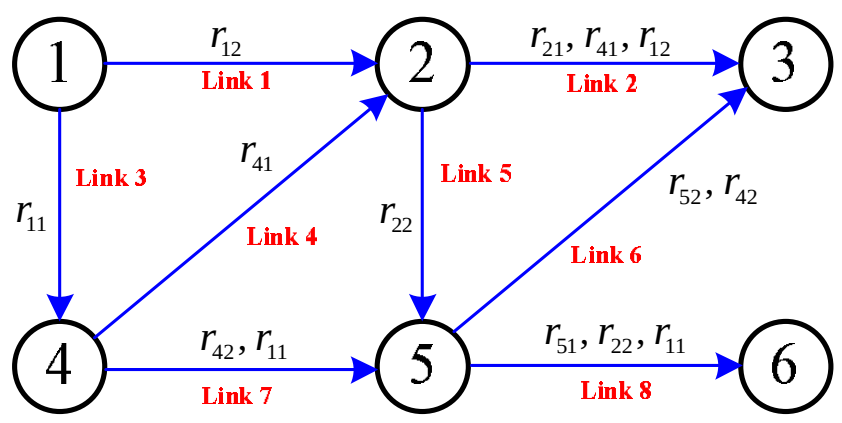

Fig. 1. A specific example of the multihop wireless network with $K = 6$ nodes and $L = 8$ directed links, where each of the source nodes (i.e., node 1, 2, 4, and 5) has two data flows. Destinations of the data flows and the routing information are summarized in the Table shown in Fig. 2. If a node has multiple outgoing links (e.g., node 1), we assume these links (e.g., link 1 and link 3) occupy different subbands and do not interfere with each other.

| Sources | | Links #1 | #2 | #3 | #4 | #5 | #6 | #7 | #8 | Destinations |
|---|---|---|---|---|---|---|---|---|---|---|
| Node 1 | Flow 1: $x_{11}$ | 0 | 0 | 1 | 0 | 0 | 0 | 1 | 1 | Node 6 |
| | Flow 2: $x_{12}$ | 1 | 1 | 0 | 0 | 0 | 0 | 0 | 0 | Node 3 |
| Node 2 | Flow 1: $x_{21}$ | 0 | 1 | 0 | 0 | 0 | 0 | 0 | 0 | Node 3 |
| | Flow 2: $x_{22}$ | 0 | 0 | 0 | 0 | 1 | 0 | 0 | 1 | Node 6 |
| Node 4 | Flow 1: $x_{41}$ | 0 | 1 | 0 | 1 | 0 | 0 | 0 | 0 | Node 3 |
| | Flow 2: $x_{42}$ | 0 | 0 | 0 | 0 | 0 | 1 | 1 | 0 | Node 3 |
| Node 5 | Flow 1: $x_{51}$ | 0 | 0 | 0 | 0 | 0 | 0 | 0 | 1 | Node 6 |
| | Flow 2: $x_{52}$ | 0 | 0 | 0 | 0 | 0 | 1 | 0 | 0 | Node 3 |

Fig. 2. Summary of the destinations of data flows and routing information of the example network shown in Fig. 1.

In our problem, we consider the system to be operating in $N_F > 1$ independent subbands between any pair of nodes. For each link, it sees $N_F$ channel gain coefficients and we denote the CSI as $\mathbf{h}_l(t) \triangleq [h_{l1}(t)\ \ h_{l2}\ \ \cdots\ \ h_{lN_F}(t)]^T$. We denote the power allocation to the $l^{th}$ link as $\mathbf{P}_l = [P_{l1}\ P_{l2}\ \cdots\ P_{lN_F}]^T$, where $P_{ln}$ is the transmit power of link $l$ on the $n^{th}$ subband. Furthermore, to deal with the interference across links, we apply *multiuser detection* (MUD) at each receiving node to recover the data from all the interfered links by successively canceling the interference. This technique can be applied as long as the transmitting rate for each link lies in the capacity region $C_{kn}$[1]. Let $\mathcal{I}_k$ denote the set of interfered links at the receiving node $k$. For example, $\mathcal{I}_2 = \{1, 4\}$, $\mathcal{I}_3 = \{2, 6\}$ in Fig. 1. Then the capacity region $C_{kn}(\mathbf{P}; \mathbf{h}(t))$ seen by the receiving node $k$ on the $n^{th}$ subband at time-slot $t$ is given by [21] ($\forall k \in \mathcal{K}, 1 \le n \le N_F$)

$$\left\{\mathbf{c}_{kn} : \sum_{j\in\mathcal{J}} c_{jn} \le \log\left(1 + \frac{\sum_{\{j\in\mathcal{J}\}} |h_{jn}(t)|^2 P_{jn}}{\sigma^2}\right), \forall\mathcal{J} \subseteq \mathcal{I}_k\right\} \tag{5}$$

where the vector $\mathbf{c}_{kn} \triangleq \left[c_{1n}\ \ c_{2n}\ \ \cdots\ \ c_{|\mathcal{I}_k|n}\right]^T$ denotes the achievable data rate vector of all the interfered links at the $n^{th}$ subband at receiving node $k$, and $\sigma^2$ denotes the noise variance. For example, $\mathbf{c}_{2n} = [c_{1n}\ \ c_{4n}]^T$, $\mathbf{c}_{3n} = [c_{2n}\ \ c_{6n}]^T$ in Fig. 1. The MCFC-MCPA problem can be formulated as follows.

*Problem 2 (MCFC-MCPA under Time-varying CSI):* Suppose that source node $k$ has $R_k$ commodities (i.e., $R_k$ different data flows), the MCFC-MCPA problem is given by

$$\underset{\{\mathbf{r}_k\succeq\mathbf{0}\},\{\mathbf{P}_l\succeq\mathbf{0}\}}{\text{maximize}} \quad \sum_{k=1}^{K}\sum_{s=1}^{R_k} f_{ks}(r_{ks}) \tag{6}$$

$$\text{subject to} \quad \sum_{(k,s)\in\mathcal{R}_l} r_{ks} \le \sum_{n=1}^{N_F} c_{ln},\ \forall l \in \mathcal{E}, \tag{7}$$

$$\mathbf{c}_{kn} \in C_{kn}(\mathbf{P}; \mathbf{h}(t)),\ \forall k \in \mathcal{K}, n \le N_F, \tag{8}$$

$$\sum_{l\in\mathcal{P}_k} \mathbf{1}^T\mathbf{P}_l \le P_{k,max},\ \forall k \in \mathcal{K}, \tag{9}$$

where $\mathbf{r}_k \triangleq [r_{k1}\ \ r_{k2}\ \ \cdots\ \ r_{kR_k}]^T \succeq \mathbf{0}$ is the vector of data flows of the $R_k$ commodities at source node $k$; $\mathbf{c} \triangleq \{c_{ln}, \forall l \in \mathcal{E}, 1 \le n \le N_F\}$ is the collection of all the auxiliary variables which are the achievable rates of the $l^{th}$ link at the $n^{th}$ subband; $\mathcal{P}_k$ denotes the set of *outgoing* links[2] at the $k^{th}$ node; $P_{k,max}$ denotes the per-node sum-power constraint.

Note that *Problem 2* is a special case of *Problem 1*. Specifically, the optimization variables $\{\mathbf{x}_k\}_{k=1}^K$ in *Problem 1* correspond to the data rate control variables $\{\mathbf{r}_k\}_{k=1}^K$, power control variables $\{\mathbf{P}_l\}_{l=1}^L$ and the auxiliary variables $\{c_{ln}, \forall l \in \mathcal{E}, 1 \le n \le N_F\}$ in *Problem 2*, with $M_k = R_k + 2N_F|\mathcal{P}_k|$. Similarly, the concave objective functions $\{f_k(\mathbf{x}_k; \mathbf{h}(t))\}_{k=1}^K$ in *Problem 1* correspond to $\left\{\sum_{s=1}^{R_k} f_{ks}(r_{ks})\right\}_{k=1}^K$ in *Problem 2*. The concave constraint functions $\{g_i(\mathbf{x}; \mathbf{h}(t))\}_{i=1}^{N_c}$ in *Problem 1* correspond to the constraints specified by (7), (8) and (9) in *Problem 2*, with $N_c = K + L + \sum_{k=1}^K \left(2^{|\mathcal{I}_k|} - 1\right)$. Observe that the capacity region given by the above formulation is a convex simplex and thanks to the auxiliary variable $c_{ln}$, which denotes the feasible rate at link $l$ and the $n^{th}$ subband, the constraints are strictly convex. Hence the optimization problem is strictly convex, provided that we have a set of strictly concave utility functions $f_{ks}$.

### *D. Primal-Dual Iterative Solution for Quasi-Static CSI*

In the existing literature, Lagrangian primal-dual algorithms are widely used for solving convex constrained optimization problems. Specifically, the Lagrangian of *Problem 2* is given in (10), where $\{\mathcal{J}_{k,i}\}_{i=1}^{2^{|\mathcal{I}_k|}-1}$ denote all the nonempty subsets of $\mathcal{I}_k$, the collection of all the forwarding links to node $k$. Let vectors $\mathbf{x} = [\mathbf{P}\ \ \mathbf{r}\ \ \mathbf{c}]^T$ and $\boldsymbol{\lambda} = [\lambda^{(\mathbf{P})}\ \ \lambda^{(\mathbf{r})}\ \ \lambda^{(\mathbf{MUD})}]^T$ denote the collection of primal variables and dual variables respectively. The primal-dual gradient algorithm is given by

$$\mathbf{x}(t+1) = [\mathbf{x}(t) + \alpha\nabla_{\mathbf{x}}\mathcal{L}(\mathbf{x}(t), \boldsymbol{\lambda}(t); \mathbf{h}(t))]^+, \tag{11}$$

$$\boldsymbol{\lambda}(t+1) = [\boldsymbol{\lambda}(t) - \alpha\nabla_{\boldsymbol{\lambda}}\mathcal{L}(\mathbf{x}(t), \boldsymbol{\lambda}(t); \mathbf{h}(t))]^+, \tag{12}$$

where $\alpha$ denotes the stepsize; the projection operation, namely $[\mathbf{a}]^+ = \max\{\mathbf{a}, \mathbf{0}\}$, shall be understood componentwisely; the vectors $\nabla_{\mathbf{x}}\mathcal{L}(\mathbf{x}(t), \boldsymbol{\lambda}(t); \mathbf{h}(t))$ and $\nabla_{\boldsymbol{\lambda}}\mathcal{L}(\mathbf{x}(t), \boldsymbol{\lambda}(t); \mathbf{h}(t))$ denote the gradients of the Lagrangian $\mathcal{L}(\cdot, \cdot; \mathbf{h}(t))$ w.r.t. $\mathbf{x}$ and $\boldsymbol{\lambda}$, respectively. It has been shown that $(\mathbf{x}(t), \lambda(t))$ in (11)-(12) converges to $(\mathbf{x}^*, \boldsymbol{\lambda}^*)$ as $t \to \infty$ under quasi-static GCSI for a strictly convex problem [8]–[10], [20].

### *E. Distributive Implementation Considerations*

Note that, in *Problem 2*, although the objective function has a decomposable structure, the constraints, such as the flow balance constraints (7), the link constraints (8) and the power constraints (9), are coupled among nodes. As a result, the computations of iterative updates (11)-(12) require global coordinates. For distributive implementation, dual decomposition methods are commonly used [1]–[4] to allocate the coupled resources by pricing, in which the price levels (Lagrangian Multipliers (LM)) control the resources allocated to each subproblem and the price levels (LMs) are updated by a master problem. By these methods, the original problem (6) can be decomposed into three separate subproblems,

$$g_I(\mathbf{c}, \lambda) = \max_{\{\mathbf{r}\succeq\mathbf{0}\}} \sum_{k=1}^K\sum_{s=1}^{R_k} f_{ks}(r_{ks}) - \sum_{(k,\,s)}\sum_{l:(k,s)\in\mathcal{R}_l} \lambda_l^{(r)} r_{ks}$$

$$g_{II}(\lambda) = \max_{\{\mathbf{c}\succeq\mathbf{0}\}} \sum_{l\in\mathcal{E}}\sum_{n=1}^{N_F} c_{ln}\left(\lambda_l^{(r)} - \sum_{k:\,k\in\mathcal{R}_l}\sum_{i:\,l\in\mathcal{J}_{k,i}} \lambda_{kni}^{(\text{MUD})}\right)$$

and

$$g_{III}(\lambda) = \max_{\{\mathbf{P}\succeq\mathbf{0}\}} \sum_{k\in\mathcal{K}}\sum_{n=1}^{N_F}\sum_{i-1}^{2^{|\mathcal{I}_k|}-1} \lambda_{kni}^{(\text{MUD})} \log\left(1 + \frac{\sum_{j\in\mathcal{J}_i}|h_{jn}|^2 P_{jn}}{\sigma^2}\right)$$

$$\text{s.t.} \quad \sum_{l\in\mathcal{P}_k} \mathbf{1}^T\mathbf{P}_l \le P_{k,max}\ \forall k \in \mathcal{K}$$

[1] For example, we assume there is MUD at the receiving node 2 so that Link 1 and Link 4 do not interfere with each other as shown in Fig. 1.

[2] All the outgoing links operate on different subbands. For example, we have $\mathcal{P}_1 = \{1, 3\}$ and $\mathcal{P}_4 = 4, 7$ in Fig. 1.

$$\mathcal{L}(\mathbf{P},\,\mathbf{r},\,\mathbf{c},\,\lambda)=\sum_{k=1}^{K}\sum_{s=1}^{R_k}f_{ks}(r_{ks})$$
$$+\sum_{l\in\mathcal{E}}\lambda_l^{(r)}\left(\sum_{n=1}^{N_F}c_{ln}-\sum_{(k,\,s)\in\mathcal{R}_l}r_{ks}\right)+\sum_{k=1}^{K}\sum_{n=1}^{N_F}\sum_{i=1}^{2^{|\mathcal{I}_k|}-1}\lambda_{kni}^{(\mathrm{MUD})}\left(\log\left(1+\frac{\sum_{j\in\mathcal{J}_{k,i}}|h_{jn}|^2P_{jn}}{\sigma^2}\right)-\sum_{j\in\mathcal{J}_{k,i}}c_{jn}\right)\quad(10)$$

The master problem is to minimize $g_I(\mathbf{c},\,\lambda)+g_{II}(\lambda)+g_{III}(\lambda)$ with respect to auxiliary variables $\{\mathbf{c}\succeq\mathbf{0}\}$ and dual variables $\{\lambda\succeq\mathbf{0}\}$. Note that given the data rates $\{\mathbf{c}\}$ and the prices $\{\lambda\}$, the maximization problem in $g_I(\mathbf{c},\,\lambda)$ is decoupled into each source commodity $(k,\,s)$, i.e. it can be computed distributively in each source node $k$. Note that to avoid trivial solutions in $g_{II}(\lambda)$, we require $\lambda_l^{(r)}=\sum_{k:\,k\in\mathcal{R}_l}\sum_{i:\,l\in\mathcal{J}_{k,i}}\lambda_{kni}^{(\mathrm{MUD})}$. For $g_{III}(\lambda)$, since it still has coupling power constriants, we apply a further dual decomposition and obtain second level subproblems as:

$$g_{IV}(\lambda)=\max_{\{\mathbf{P}\succeq\mathbf{0}\}}\sum_{k\in\mathcal{K}}\sum_{n=1}^{N_F}\sum_{i=1}^{2^{|\mathcal{I}_k|}-1}\lambda_{kni}^{(\mathrm{MUD})}\log\left(1+\frac{\sum_{j\in\mathcal{J}_i}|h_{jn}|^2P_{jn}}{\sigma^2}\right)-\sum_{k\in\mathcal{K}}\lambda_k^{(P)}\sum_{l\in\mathcal{P}_k}\mathbf{1}^T\mathbf{P}_l\quad(13)$$

One remark is that the objective function in the subproblem (13) is coupled by the interfering multi-access links towards the receiving nodes. However, the subproblem (13) can still be solved locally at each receiving node running MUD and hence, the subproblem (13) can be solved locally.

The following summarizes the distributive iterative solution for solving the MCFC-MCPA problem in (6) based on the above decompositions. Here we ignore the algorithm initialization in which all the parameters can take arbitrary values. At each time slot $t$, we implement the following updates simultaneously.

- At each source node $k$, based on the received message $\lambda_l^{(r)}$ the commodity rate $r_{ks}$ is updated by locally solving
$$\max_{r_{ks}\geq 0}f_{ks}(r_{ks})-\sum_{l:(k,s)\in\mathcal{R}_l}\lambda_l^{(r)}r_{ks}$$
- Each link $l$ calculates and broadcasts $c_{ln}$and $\lambda_l^{(r)}$ by the received message of $\lambda_{kni}^{(\mathrm{MUD})}$ as
$$c_{ln}(t+1)=\left[c_{ln}(t)+\alpha_l\left(\lambda_l^{(r)}-\sum_{k,n,i}\lambda_{kni}^{(\mathrm{MUD})}\right)\right]^+$$
$$\lambda_l^{(r)}(t+1)=\left[\lambda_l^{(r)}(t)-\gamma_l\left(\sum_{n=1}^{N_F}c_{ln}-\sum_{(k,\,s)\in\mathcal{R}_l}r_{ks}\right)\right]^+$$
- Each receiving node $k$ updates the power $P_{jn}$ to the corresponding transmitter $j$ by iterating one step to solve (13) as
$$P_{jn}(t+1)=\left[P_{jn}(t)+\alpha_p\left(\sum_i\lambda_{kni}^{(\mathrm{MUD})}\frac{|h_{jn}|^2}{\sigma^2+\sum_j|h_{jn}|^2}-\lambda_k^{(P)}\right)\right]^+$$
$\lambda_{kni}^{(\mathrm{MUD})}$ is also updated by minimizing the master problem with one step in the iteration
$$\lambda_{kni}^{(\mathrm{MUD})}(t+1)=\left[\lambda_{kni}^{(\mathrm{MUD})}(t)-\gamma_{\mathrm{M}}\left(\log\left(1+\frac{\sum_j|h_{jn}|^2P_{jn}}{\sigma^2}\right)-\sum_j c_{jn}\right)\right]^+$$
where $\alpha_p,\,\gamma_{\mathrm{M}}>0$ are constants and $\lambda_{kni}^{(\mathrm{MUD})}$ together with $\lambda_k^{(P)}$ are quantified by message passing.
- Each transmit node $j$ updates LM $\lambda_k^{(P)}$ based on messages $P_{jn}$ from the receiving node with constant value $\gamma_p$ by
$$\lambda_k^{(P)}(t+1)=\left[\lambda_k^{(P)}(t)-\gamma_p\left(P_{k,max}-\sum_{l\in\mathcal{P}_k}\mathbf{1}^T\mathbf{P}_l\right)\right]^+$$

In the existing literature [2]–[4], the channel coefficients $\{\mathbf{h}_{ln}\}$ (channel state information CSI) are assumed to be quasi-static and there is a long enough time for the algorithm to converge before the CSI changes. However, in practice, this quasi-static assumption is difficult to achieve and the algorithm may only be able to iterate very few times before the CSI changes, for explicit message passing (such as the Lagrangian multipliers $\lambda_l^{(r)}$ and $\lambda_k^{(P)}$) between nodes are involved in the solution above. While the traditional notion of convergence discusses whether an iterative algorithm will converge to the *targets* ($\mathbf{P}_l^*$ and $r_{ks}^*$), we need to extend the notion of convergence when dealing with time varying CSI because the targets $\mathbf{P}_l(\mathbf{h}(t))^*$ and $r_{ks}(\mathbf{h}(t))^*$ become *moving targets*. Furthermore, the time varying CSI also causes a subtle problem in that the constraints of the problem (such as flow balance constraints, link constraints and power constraints) may not be satisfied with $\mathbf{P}_l(\mathbf{h}(t))^*$, $\mathbf{c}_{ln}(\mathbf{h}(t))^*$ and $r_{ks}(\mathbf{h}(t))^*$ at every time slot and this refers to the *constraint outage*. The issues of *convergence w.r.t. moving targets* and *constraint outage* will be addressed in the following sections.

## III. Randomly Switched Modeling and Convergence Behavior Analysis

In general, one could use the (traditional) primal-dual gradient update in (11) and (12) to solve the NUM problem for quasi-static CSI. The transient behavior of the primal-dual update equations in (11) and (12) can be characterized by an *algorithm trajectory* of an associated nonlinear dynamic system and the optimal solution is the *equilibrium point* of the nonlinear system. Since the optimal solution $(\mathbf{x}^*(\mathbf{h}(t)), \boldsymbol{\lambda}^*(\mathbf{h}(t)))$ depends on the CSI, time-varying CSI corresponds to a randomly moving optimal solution (or a randomly moving equilibrium point) and the convergence behavior of the algorithm can be measured by how well the algorithm trajectory could track the moving equilibrium point. In this section, we shall first generalize the basic primal-dual update algorithm in (11) and (12) for the general NUM problem (1) under time-varying CSI. We shall then utilize the *randomly switched system* nonlinear control theory to analyze the convergence behavior of the generalized algorithm.

### A. Primal-Dual Scaled Gradient Algorithm under Time-Varying CSI

In this section, we generalize the Arrow-Hurwicz-Uzawa type primal-dual gradient method [8]–[10] for computing the saddle point of the Lagrangian $\mathcal{L}(\mathbf{x}, \boldsymbol{\lambda}; \mathbf{h}(t))$ under time-varying channel conditions. In particular, scaling matrices $\mathbf{D}_{\mathbf{x}}(t)$ and $\Lambda(t)$ are introduced into the primal-dual iterations as follow:

$$\mathbf{x}\left(t+\overline{T}\right) = \left[\mathbf{x}(t) + \mathbf{D}_{\mathbf{x}}(t)\nabla_{\mathbf{x}}\mathcal{L}\left(\mathbf{x}(t), \boldsymbol{\lambda}(t); \mathbf{h}\left(t+\overline{T}\right)\right)\right]^+ \quad (14)$$

$$\boldsymbol{\lambda}\left(t+\overline{T}\right) = \left[\boldsymbol{\lambda}(t) - \boldsymbol{\Lambda}(t)\nabla_{\boldsymbol{\lambda}}\mathcal{L}\left(\mathbf{x}(t), \boldsymbol{\lambda}(t); \mathbf{h}\left(t+\overline{T}\right)\right)\right]^+ \quad (15)$$

where the integer constant $\overline{T} \geq 1$ denotes the *period* (in terms of time-slots) of updating the resource allocation vectors $\mathbf{x}(t)$ as well as the LM $\boldsymbol{\lambda}(t)$; the symmetric positive definite matrix $\mathbf{D}_{\mathbf{x}}(t) \in \mathbb{C}^{M_k \times M_k}$ and $\boldsymbol{\Lambda}(t) \in \mathbb{C}^{N_c \times N_c}$ are the *scaling matrices*, which are introduced to speed up the convergence rate under time-varying CSI; the vectors $\nabla_{\mathbf{x}}\mathcal{L}\left(\mathbf{x}(t), \boldsymbol{\lambda}(t); \mathbf{h}\left(t+\overline{T}\right)\right)$ and $\nabla_{\boldsymbol{\lambda}}\mathcal{L}\left(\mathbf{x}(t), \boldsymbol{\lambda}(t); \mathbf{h}(t+\overline{T})\right)$ denote the gradients of the Lagrangian $\mathcal{L}\left(\cdot, \cdot; \mathbf{h}\left(t+\overline{T}\right)\right)$ w.r.t. $\mathbf{x}(t)$ and $\boldsymbol{\lambda}(t)$, respectively. The iterative algorithm in (14) and (15) is called the PDSGA.

Note that the concept of using the scaling matrix to speed up the convergence rate of iterative algorithms in quasi-static fading channels has appeared in previous literature. For example, when $\mathbf{D}_{\mathbf{x}}(t) = diag\{\alpha\}$ and $\boldsymbol{\Lambda}(t) = diag\{\alpha\}$, the iterative algorithm in (14) and (15) reduces to the standard Arrow-Hurwicz-Uzawa algorithm [8]–[10]. On the other hand, in [1], [22], the scaling matrices are chosen to be the diagonal entries of the inverse Hessian. In [20], [23], the scaling matrices are chosen to be the inverse Hessian matrix. In this section, we shall focus on analyzing the tracking performance of the PDSGA under time-varying CSI for general scaling matrices. In Section IV, we shall propose a low complexity algorithm to dynamically adjust the scaling matrix according to the CSI.

*Remark 1: (Tradeoff between Message Passing Overhead and Performance)* In the iteration process (14) and (15), the network nodes update their resource control vectors *every* $\overline{T}$ *time-slots*. This parameter $\overline{T}$ controls the underlying tradeoff between message passing overhead and the performance. For example, when the *updating period* $\overline{T} = 1$, new resource control results will be used in each time-slot at the expense of high message passing overhead. On the other hand, when $\overline{T}$ is large, the same resource control vector $\mathbf{x}(t)$ will be applied for $\overline{T}$ time-slots before the next update at time-slot $\left(t+\overline{T}\right)$ and hence, this corresponds to low message passing overhead at the expense of some performance loss.

*Remark 2 (Impact of the Time-Varying GCSI):* As the FSMC $\{\mathbf{h}(t)\}$ jumps from one state to another randomly, the saddle point $\left(\mathbf{x}^*(\mathbf{h}(t)),\ \boldsymbol{\lambda}^*(\mathbf{h}(t))\right)$ of the Lagrangian $\mathcal{L}(\mathbf{x}, \boldsymbol{\lambda}; \mathbf{h}(t))$ also changes with time randomly. As a result, the primal-dual scaled gradient iteration process (14) and (15) would not converge to a single point but rather *track* the moving saddle point $(\mathbf{x}^*(\mathbf{h}(t)), \boldsymbol{\lambda}^*(\mathbf{h}(t)))$. The *convergence property* and *tracking performance* of the PDSGA in (14) and (15) are the focus of this paper.

### B. Randomly Switched System Modeling

Randomly switched systems are *piecewise deterministic* stochastic systems, i.e., between any two consecutive *switching instants*, the dynamics are deterministic [13]. Formally, a *discrete-time randomly switched system* is defined as follows [13].

*Definition 1 (Discrete-time Randomly Switched System):* A discrete-time *randomly switched system* consists of a family of *subsystems*, and a random *switching signal* that specifies the *active* subsystem at every time-slot. Mathematically,

$$\mathbf{y}(t+\overline{T}) = \mathscr{F}_u(\mathbf{y}(t)), \text{when } \tau(t) = u \in \mathcal{U}, \quad (16)$$

where $\mathbf{y}(t)$ denotes the system state; $\mathscr{F}_u(\mathbf{y}(t))$ denotes the $u^{th}$ *subsystem*; and $\tau(n)$ is the switching signal with state space $\mathcal{U}$. ■

For the FSMC $\{\mathbf{h}(t)\}$, the channel fading process stays in a state $\mathbf{h}^{(q)}$ for a *random sojourn time* [3] of $T_q$ time-slots, and then jumps to another state randomly. During the $T_q$ time-slots, the channel fading coefficients $\{\mathbf{h}(t)\}$ remain constant (i.e., $\mathbf{h}(t) = \mathbf{h}^{(q)}$) and the system is deterministic. We thus can model the dynamics of the proposed primal-dual scaled gradient iteration process (14) and (15) as a *randomly switched system*, with the FSMC $\{\mathbf{h}(t)\}$ being the *switching signal* and the channel state $\mathbf{h}^{(q)}$ corresponding to the $q^{th}$ *subsystem* [13], for all $q \in \mathcal{Q}$.

To obtain the dynamics of the randomly switched system of the proposed primal-dual scaled gradient iteration process (14) and (15) explicitly, we rewrite iterations given by equation (14) and (15) into one vector form, i.e.,

$$\mathbf{y}\left(t+\overline{T}\right) = \left[\mathbf{y}(t) + \mathbf{D}(t)\nabla\mathcal{L}\left(\mathbf{y}(t); \mathbf{h}\left(t+\overline{T}\right)\right)\right]^+, \quad (17)$$

where $\mathbf{y}(t) = [\mathbf{x}(t)\ \boldsymbol{\lambda}(t)]^T \in \mathbb{R}_+^{(M+N_c)}$ is the collection of all the iterates, $\mathbf{D}(t) = blkdgl\{\mathbf{D}_{\mathbf{x}}(t),\ \boldsymbol{\Lambda}(t)\}$ is the scaling matrix for the vector $\mathbf{y}(t)$, and $\nabla\mathcal{L}\left(\mathbf{y}(t); \mathbf{h}\left(t+\overline{T}\right)\right)$ denotes the overall *fictitious gradient*[4] of the Lagrangian $\mathcal{L}(\mathbf{x}, \boldsymbol{\lambda};$

[3]Sojourn time for a switched system is the total time a subsystem spends in a state before leaving it.

[4]Note that $\nabla\mathcal{L}(t)$ defined in (18) is the search direction in the primal-dual gradient algorithm. It is, however, not the actual gradient of the Lagrangian function $\mathcal{L}$.

$\mathbf{h}\left(t+\overline{T}\right)$ ), evaluated at $(\mathbf{x}(t), \boldsymbol{\lambda}(t))$, i.e.,

$$\nabla\mathcal{L}\left(\mathbf{y}(t);\mathbf{h}\left(t+\overline{T}\right)\right) = \begin{bmatrix} \nabla_{\mathbf{x}}\mathcal{L}\left(\mathbf{x}(t),\boldsymbol{\lambda}(t);\mathbf{h}\left(t+\overline{T}\right)\right) \\ -\nabla_{\boldsymbol{\lambda}}\mathcal{L}\left(\mathbf{x}(t),\boldsymbol{\lambda}(t);\mathbf{h}(t+\overline{T}\right) \end{bmatrix} \triangleq \nabla\mathcal{L}\left(t\right). \quad (18)$$

Moreover, we denote the *equilibrium point* of the $q^{th}$ subsystem as $\mathbf{y}_q^* = \left(\mathbf{x}_q^*, \boldsymbol{\lambda}_q^*\right)$, $q \in \mathcal{Q}$. As a result, we end up with a randomly switched system with $Q$ subsystems, whose dynamics is given by (17) and the $Q$ equilibrium points are given by $\left\{\mathbf{y}_q^*\right\}_{q=1}^{Q}$. For notational convenience, $\nabla\mathcal{L}\left(\mathbf{x}(t),\boldsymbol{\lambda}(t);\mathbf{h}\left(t+\overline{T}\right)\right)$ will be simplified as $\nabla\mathcal{L}\left(t\right)$ hereafter.

*Remark 3 (Interpretation of the Equilibrium Points):* Note that, an equilibrium point of the dynamical system defined in (17) is a saddle point of the Lagrangian $\mathcal{L}\left(\mathbf{x},\boldsymbol{\lambda};\mathbf{h}(t)\right)$ defined in (2). Moreover, for each equilibrium point $\mathbf{y}_q^* = \left(\mathbf{x}_q^*, \boldsymbol{\lambda}_q^*\right)$, the first component $\mathbf{x}_q^*$ is the unique global optimal solution of the NUM problem (1), when the channel fading process $\{\mathbf{h}(t)\}$ is in state $\mathbf{h}^{(q)}$, $\forall q \in \mathcal{Q}$.

### C. Convergence Analysis of the PDSGA

The *region stability* is a widely adopted performance measure of iterative algorithms in time-varying environments, especially for randomly switched and hybrid systems [11], [14], [24]. When the CSI is time-varying (e.g., the FSMC model), the equilibrium point of the network is also changing and hence, the *algorithm trajectory* of the primal-dual scaled gradient algorithm will not converge to a single point but rather a *limit region* as illustrated in Fig. 3 for the one dimensional case. We shall formally define *region stability* as follows.

*Definition 2: (Region Stability of Randomly Switched Systems)* A discrete-time randomly switched system with state vector $\mathbf{y}(t)$ is said to be stable w.r.t. a *limit region* $\mathcal{Y}$, if for every trajectory $\mathbf{y}\left(t,\mathbf{y}(0)\right)$, there exists a point of time $T_0\left(\mathbf{y}(0)\right)$ such that from then on, the trajectory is always in the limit region $\mathcal{E}$. Mathematically, $\forall\ \mathbf{y}\left(t,\mathbf{y}(0)\right)$, $\exists\ T_0\left(\mathbf{y}(0)\right)$, such that $\mathbf{y}\left(t,\mathbf{y}(0)\right) \in \mathcal{Y}, \forall\ t \geq T_0\left(\mathbf{y}(0)\right)$. ■

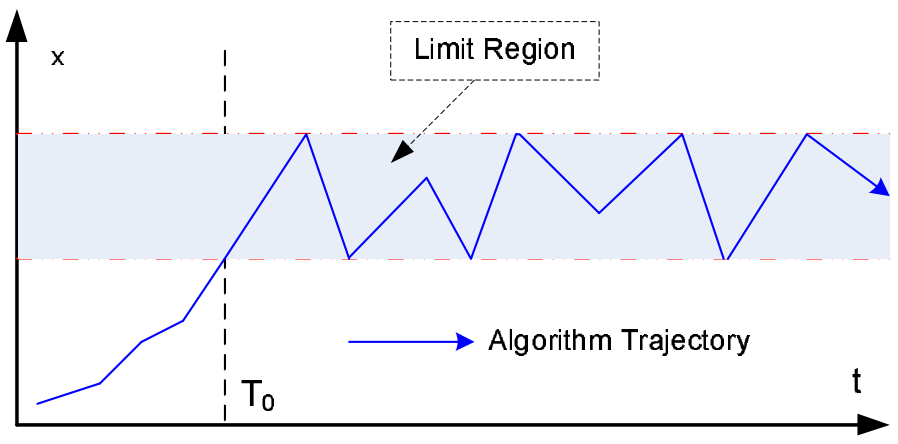

Fig. 3. An illustration of the region stability of a randomly switched system with one-dimensional state space [24]. From time-slot $T_0$ and onwards, the trajectory goes outside the limit region with probability $\mathrm{P}_{\bar{\mathcal{Y}}}$ given in equation (21).

For analyzing the convergence behavior of the PDSGA in (17), we shall first consider the case when the channel fading process $\{\mathbf{h}(t)\}$ is in a particular channel state $\mathbf{h}^{(q)}$ for a sojourn time of $T_q$ time-slot, $\forall q \in \mathcal{Q}$. Here the sojourn time $T_q$ is defined to be the time period in which the channel state $\mathbf{h}^{(q)}$ remains unchanged. Let $\mathbf{y}_q^* = \left(\mathbf{x}_q^*, \boldsymbol{\lambda}_q^*\right)$ denote the equilibrium point of the iteration (17) when the channel fading process $\{\mathbf{h}(t)\}$ is in state $\mathbf{h}^{(q)}$, $\forall q \in \mathcal{Q}$. We first introduce the following lemma, which summarizes the *contraction property* of the PDSGA in (17).

*Lemma 1 (Contraction Property of the PDSGA):* Suppose that $\mathbf{D}\left(\mathbf{h}^{(q)}\right)$ is designed in a way such that

$$\beta\left(\mathbf{D}(t)\right) \triangleq \max_{\bar{\mathbf{y}}} \left\|\mathbf{I} + \mathbf{D}(t)\nabla^2\mathcal{L}\left(\bar{\mathbf{y}};\mathbf{h}^{(q)}\right)\right\|_2 \leq \beta_q < 1, \quad (19)$$

where $\bar{\mathbf{y}} \in \mathbb{R}_+^{(M+N_c)}$ is the collection of primal and dual variables. For one iterative update under the same channel state $\mathbf{h}^{(q)}$, we have

$$\left\|\mathbf{y}\left(t+\overline{T}\right) - \mathbf{y}_q^*\right\|_2 \leq \beta_q \left\|\mathbf{y}(t) - \mathbf{y}_q^*\right\|_2, \quad (20)$$

where $\nabla^2\mathcal{L}\left(\bar{\mathbf{y}};\mathbf{h}^{(q)}\right) \in \mathbb{R}_+^{(M+N_c)\times(M+N_c)}$ denotes the *second-order* derivative of the Lagrangian $\mathcal{L}\left(\mathbf{y}(t);\mathbf{h}^{(q)}\right)$ w.r.t. $\mathbf{y}$, evaluated at $\bar{\mathbf{y}}$; $\beta_q$ is referred to as the *contraction modulus* at the $q^{th}$ channel state, $\forall q \in \mathcal{Q}$; and the *matrix-2 norm* is defined as $\|\mathbf{A}\|_2 = \max_{\{\mathbf{x}:\ \|\mathbf{x}\|_2=1\}} \|\mathbf{A}\mathbf{x}\|_2$, with $\|\cdot\|_2$ denoting the standard Euclidean norm.

*Proof:* Please refer to Appendix A for the proof. ■

*Remark 4 (Geometric Convergence of the PDSGA):* Note that the *contraction modulus* $\beta_q$ depends on the scaling matrix $\mathbf{D}(t)$, which may exceed 1 for some inappropriate choice of $\mathbf{D}$. Hence, we shall restrict the domain of $\mathbf{D}$ such that $\beta_q < 1$. Table I summarizes the worst-case $\beta$ under the adaptive scaling matrix (in Section IV) for the 6-node joint network flow and power control problem as illustrated in Fig. 1 and Fig. 2. Observed from Table I, $\beta_q$ is always less than 1 for all $q$ and hence, there exists at least one scaling matrix control policy that satisfies (19).

Moreover, since the sojourn time $T_q$ of the channel state $\mathbf{h}^{(q)}$ is a random variable (which is geometrically distributed with parameter $\nu$), $\forall q \in \mathcal{Q}$ , and the network nodes update their resource allocation vectors $\mathbf{x}(t)$ and the LM $\boldsymbol{\lambda}(t)$ every $\overline{T}$ time-slots, there may be several updates, or only one update, or even *no* update during the $T_q$ time-slots. The number of updates $N_q$ of the PDSGA in (17) during the sojourn time $T_q$ of the channel state $\mathbf{h}^{(q)}$, $\forall q \in \mathcal{Q}$, is given by $N_q = \left\lfloor \frac{T_q}{\overline{T}} \right\rfloor$, $\forall q \in \mathcal{Q}$, where $\lfloor\cdot\rfloor$ denotes the floor function.

Using *Lemma 1*, we can derive the *region stability* of the PDSGA in (17), which is summarized in the following theorem.

*Theorem 1 (Region Stability of the PDSGA):* Under the conditions that $\beta\left(\mathbf{D}(t)\right) \leq \beta_q < 1, \forall q \in \mathcal{Q}$ (see equation (19)), at steady state (i.e., for sufficiently large $t$), the probability that the iterates $\mathbf{y}(t)$ generated by (17) being inside the limit region $\mathcal{Y}$ is given by:

$$\mathrm{P}_{\mathcal{Y}} \triangleq \lim_{t\to+\infty} \Pr\left\{\mathbf{y}(t) \in \mathcal{Y}\right\} \geq 1 - \min\left\{1, \frac{\beta\overline{T}}{(1-\beta)\,\overline{N}}\right\} \quad (21)$$

where $\beta = \max_{\{\forall q\in Q\}}\left\{\beta\left(\mathbf{D}(t)\right)\right\}$ is the maximum contraction modulus of all channel states; $\overline{N} = \frac{1}{1-\nu^{L^2}}$ is the *average sojourn time* of the FSMC $\{\mathbf{h}(t)\}$; and the limit region $\mathcal{Y}$ is given by

$$\mathcal{Y} = \bigcup_{q=1}^{Q}\left\{\mathbf{y}\middle|\ \left\|\mathbf{y} - \bar{\mathbf{y}}_q^*\right\|_2 \leq \delta\right\} \quad (22)$$

where $\delta = \max_{\{\forall q,r\in\mathcal{Q},q\neq r\}} \left\|\mathbf{y}_q^* - \mathbf{y}_r^*\right\|_2$ denotes the maximum distance between two equilibrium points in set of all equilibrium points $\left\{\mathbf{y}_q^*, \forall\ q \in \mathcal{Q}\right\}$.

*Proof:* Please refer to Appendix B for the proof. ■

*Remark 5: (Region Stability versus Average Sojourn Time and Update Period)* The upper bound given in (21) in *Theorem 1* depends on the average sojourn time $\overline{N}$ of the channel fading process $\{\mathbf{h}(t)\}$ and the update period $\overline{T}$. The average sojourn time $\overline{N}$ can be thought of as an indicator of the channel fading rate. The larger the $\overline{N}$ is, the channel changes more slowly and the larger the $\mathrm{P}_{\mathcal{Y}}$ is. On the other hand, the update period $\overline{T}$ represents the tradeoff between the message passing overhead and the tracking performance of the PDSGA in (17). The smaller the update period $\overline{T}$ is, the greater the message passing overhead becomes and the larger the $\mathrm{P}_{\mathcal{Y}}$ is. In particular, we have $\mathrm{P}_{\mathcal{Y}} = 1 - \mathcal{O}\left(\overline{T}/\overline{N}\right)$.

### D. Asymptotic Analysis of the Tracking Errors

In this section, we shall derive the closed-form expressions of the asymptotic order of growth of the tracking errors associated with the PDSGA in (17) under the FSMC channel model. Here, we shall study both the expected-absolute-error (EAE) and the mean-square-error (MSE). We first give the formal definitions of EAE and MSE below.

*Definition 3 (EAE and MSE):* The EAE (or MSE) is defined to be the expectation of the distance (or *squared* distance) between the iterate $\mathbf{y}(t)$ and the corresponding target optimal solution $\mathbf{y}_q^*$, i.e.,

$$\begin{aligned} \mathrm{EAE}\left(\mathbf{y}(t)\right) &= \mathbb{E}_{\{\mathbf{h}(t)\}}\left\{\left\|\mathbf{y}(t) - \mathbf{y}_q^*\right\|_2\right\}, \\ \mathrm{MSE}\left(\mathbf{y}(t)\right) &= \mathbb{E}_{\{\mathbf{h}(t)\}}\left\{\left\|\mathbf{y}(t) - \mathbf{y}_q^*\right\|_2^2\right\}, \end{aligned} \quad (23)$$

where the expectation shall be taken over the stationary distribution of the FMSC $\{\mathbf{h}(t)\}$.

Note that as the switched system $\mathbf{y}(t)$ is driven by switching signal CSI $\mathbf{h}(t) = \mathbf{h}^{(q)}$, taking expectation over $\mathbf{h}(t)$ and taking expectation over subsystems $\mathbf{y}(t) = \mathbf{y}_q^*$ are equivalent in (23). Due to the randomness of the channel fading process $\{\mathbf{h}(t)\}$ and the nonlinear dynamics in (17), it is impossible to derive the exact expressions for the tracking errors. However, we can derive the asymptotic order of growth of the expected-absolute-error $\mathrm{EAE}\left(\mathbf{y}(t)\right)$ and the mean-square-error $\mathrm{MSE}\left(\mathbf{y}(t)\right)$, which are actually good enough to design the tracking error optimal scaling matrices, as shall be shown in the next section. We now summarize the main results of this section in the following theorem.

*Theorem 2: (Asymptotic Order of Growth of the EAE and the MSE)* Under the conditions that $\beta\left(\mathbf{D}(t)\right) \leq \beta_q < 1, \forall q \in \mathcal{Q}$ (see equation (19)), the asymptotic order of growth of the expected-absolute-error $\mathrm{EAE}\left(\mathbf{y}(t)\right)$ and mean-square-error $\mathrm{MSE}\left(\mathbf{y}(t)\right)$ at steady state (i.e., for sufficient large $t$) are given by:

$$\begin{aligned} \mathrm{EAE}\left(\mathbf{y}(t)\right) &= \mathcal{O}\left(\frac{\beta\overline{T}}{(1-\beta)\overline{N}}\right) \\ \mathrm{MSE}\left(\mathbf{y}(t)\right) &= \mathcal{O}\left(\frac{\beta^2\left(2\beta + \overline{T}(1-\beta)\overline{N}\right)}{(1-\beta^2)(1-\beta)\overline{N}^2}\right), \end{aligned} \quad (24)$$

where $\beta = \max_{\{\forall q\in Q\}}\left\{\beta\left(\mathbf{D}(t)\right)\right\}$ is the maximum contraction modulus of all channel states.

*Proof:* Please refer to Appendix C for the proof. ■

*Remark 6: (Tracking Errors versus Average Sojourn Time and Update Period)* The expressions of the asymptotic tracking errors EAE and MSE given in *Theorem 2* depend on the average sojourn time $\overline{N}$ of the channel fading process $\{\mathbf{h}(t)\}$ and the update period $\overline{T}$. The average sojourn time $\overline{N}$ can be thought as an indicator of the *channel fading rate*. The larger the $\overline{N}$ is, the slower the channel changes and the smaller the tracking errors are. On the other hand, the update period $\overline{T}$ represents the tradeoff between the message passing overhead and the tracking performance of the PDSGA in (17). The smaller the update period $\overline{T}$ is, the greater the message passing overhead becomes and the smaller the tracking errors are. In particular, we have $\mathrm{EAE}\left(\mathbf{y}(t)\right) = \mathcal{O}\left(\overline{T}/\overline{N}\right)$ and $\mathrm{MSE}\left(\mathbf{y}(t)\right) = \mathcal{O}\left(\overline{T}/\overline{N}\right)$.

### E. Analysis of Constraint Outage Probability and Backoff Margin

Beside the convergence error, the time varying CSI may cause constraint outage in which the constraints of the NUM problem may not be satisfied at every time slot. To quantify this undesirable penalty of time varying CSI, we shall quantify the probability of *constraint outage* in this section. Moreover, although the constraint outage is unavoidable due to the time varying CSI, we would like to minimize the probability of the constraint outage and this can be achieved by introducing a *constraint backoff margin*. Specifically, let $K$ be the constraint backoff margin corresponding to the constraint $g(\mathbf{x}) \leq 0$. The following lemma summarizes the probability of constraint outage given a constraint backoff margin.

*Lemma 2: (Probability of Constraint Outage with Backoff Margin)* Assume that there is an inequality constraint $g(\mathbf{x}) \leq 0$ in the NUM problem, where $g(\mathbf{x}) : \mathbf{x} \in \mathbb{D} \subseteq \mathbf{R}^n \mapsto \mathbf{R}$ satisfies conditions (i) $g(\mathbf{x})$ is convex and (ii) $g(\mathbf{x})$ is Lipschitz continuous on $\mathbb{D}$ with Lipschitz constant $c > 0$ such that $\forall \mathbf{x} \in \mathbb{D}$,

$$\|g(\mathbf{x}_1) - g(\mathbf{x}_2)\| \leq c\|\mathbf{x}_1 - \mathbf{x}_2\|.$$

Then, given a backoff margin $K \geq 0$, at the steady state ($t \to \infty$), the constraint outage probability for $g(\mathbf{x}) \leq 0$ is bounded by

$$\lim_{t\to\infty} \Pr\left\{g(\mathbf{x}(t)) > \epsilon\right\} \leq \min\left\{1, \frac{c\delta\beta\bar{T}}{(\epsilon + K)(1-\beta)\bar{N}}\right\} \quad (25)$$

where $\beta < 1$ is the maximum contraction modulus of all channel states for the optimization variable $\mathbf{x}(t)$, $\delta$ is the maximum distance between equilibrium points from any two adjacent channel states, $\bar{T}$ is the algorithm update period and $\bar{N}$ is the average sojourn time of the channel states.

*Proof:* Please refer to Appendix D for the proof. ■

*Remark 7: (Tradeoff between Performance and Constraint Outage)* *Lemma 2* indicates that there is a tradeoff between system performance and constraint outage via the *constraint backoff margin*. For example, when the constraint backoff margin $K$ is large, the constraint outage probability is small but the performance drops, because the margin restricts the optimization variables to stay within a shrunk optimization region and hence excludes some optimal solutions. On the other hand, the margin also reduces the outage probability that the instantaneous points from the algorithm trajectory go beyond the original optimization region (i.e. the feasible set of the problem). Note that for the same target constraint outage probability, a larger backoff margin is needed for fast changing CSI.

## IV. Tracking Error Optimization and Scaling Matrix Adaptation

In the previous section, we have established the region stability property of the PDSGA in (17) and derived the order of growth of the tracking errors. In this section, we propose a low complexity distributive algorithm for determining the dynamic scaling matrices $\{\mathbf{D}(t), \forall\, t \geq 1\}$ to minimize the tracking errors. Specifically, we shall show that the scaling matrixes shall be adaptive to the time-varying CSI $\mathbf{h}(t)$ and can be computed distributively by each small group of nodes based on the local information only.

### *A. Tracking Error Control Problem*

Using the results of the tracking error analysis in (24) as well as constraint outage probability in (25), the tracking error and the constraint outage probability are both shown to be an increasing function of the parameter $\beta = \max_{\{\forall m\}} \beta\left(\mathbf{D}(m)\right)$, where $\beta\left(\mathbf{D}(m)\right)$ denotes the contraction modulus at the $m^{th}$ stage, which is defined as

$$\beta\left(\mathbf{D}\left(m\right)\right) \triangleq \max_{\bar{\mathbf{y}}} \left\|\mathbf{I} + \mathbf{D}\left(m\right)\nabla^2\mathcal{L}\left(\bar{\mathbf{y}};\mathbf{h}(m)\right)\right\|_2, \quad (26)$$

where $\bar{\mathbf{y}} \in \mathbb{R}_+^{(M+N_c)}$ is the collection of primal and dual variables. As a result, this suggests that the dynamic scaling matrices $\mathbf{D}(m)$ should be chosen to minimize $\beta\left(\mathbf{D}(m)\right)$, which is a function of the time-varying CSI $\mathbf{h}(m)$. This is formally cast into the following optimization problem.

$$\begin{aligned} \underset{\mathbf{D}(m)}{\text{minimize}} \quad & \beta\left(\mathbf{D}(m)\right) \\ \text{subject to} \quad & \mathbf{D}(m) \succ 0, \forall m \geq 1. \end{aligned} \quad (27)$$

From (26), we can obtain the intuition that the optimal solution of (27) should be $\mathbf{D}(m) = -\nabla^2\mathcal{L}\left(\bar{\mathbf{y}}(t);\mathbf{h}(m)\right)^{-1}$. However, the computation of inverse Hessian $\nabla^2\mathcal{L}$ requires global information, which is difficult to be implemented. Furthermore, the value of $\bar{\mathbf{y}}(t)$ is not known and the solution is not attainable. To facilitate low complexity solution, we first take $\bar{\mathbf{y}} = \mathbf{y}(t)$ in (26) and $\beta\left(\mathbf{D}(m)\right)$ in (27) is replaced by:

$$\widetilde{\beta}\left(\mathbf{D}\left(t+\overline{T}\right)\right) = \left\|\mathbf{I} + \mathbf{D}\left(t+\overline{T}\right)\nabla^2\mathcal{L}\left(\mathbf{y}(t);\mathbf{h}(t)\right)\right\|_2 \quad (28)$$

where $\mathbf{y}(t)$ is the iterate generated by (17) at the $t^{th}$ time-slot. To facilitate distributive implementation of the adaptive scaling matrix, we partition the whole network into $G$ groups of nodes, and partition the collection of optimization variables $\mathbf{y}(t) = [\mathbf{y}_1(t)\,|\,\mathbf{y}_2(t)\,|\,\cdots\,|\,\mathbf{y}_G(t)]$ accordingly. We then impose a block diagonal structure to the scaling matrix $\mathbf{D}(t)$, which is given by

$$\mathbf{D}(t) = \text{blkdlg}\{\mathbf{D}_{\mathbf{y}_1}(t),\, \mathbf{D}_{\mathbf{y}_2}(t),\, \cdots,\, \mathbf{D}_{\mathbf{y}_G}(t)\} \quad (29)$$

where $\mathbf{y}_i(t)$, $\forall i = 1, \cdots, G$ is the collection of variables for the $i$-th group of nodes. A convenient choice for the scaling matrix blocks in (29) is $\mathbf{D}_{\mathbf{y}_i}(t) = -\nabla^2_{\mathbf{y}_i}\mathcal{L}(\mathbf{y}(t);\,\mathbf{h}(t))^{-1}$, which can be computed using local information within the $i^{th}$ group of nodes. Note that in wireless communication networks, nodes usually have weak connections when they are far away and hence, $\nabla^2\mathcal{L}(\mathbf{y}(t),\,\mathbf{h}(t))$ is actually a sparse matrix. Hence, imposing the block diagonal structure of $\mathbf{D}(t)$ in (28) does not cause too much performance loss as illustrated in Fig. 4-7 in the simulation section. We shall illustrate the design with an example in Section IV-B.

### *B. Example: Adaptive Scaling Matrices for the MCFC-MCPA Problem*

The Lagrangian of *Problem 2* w.r.t. the specific network topology in Fig. 1 in the MCFC-MCPA example is given by (e.g., $N_F = 2$, $R_k = 2, \forall k \in \{1,2,4,5\}$):

$$\begin{aligned} \mathcal{L}\left(\mathbf{r},\mathbf{P},\mathbf{c},\boldsymbol{\lambda}\right) &= \sum_{k\in\{1,2,4,5\}}\sum_{s=1}^{R_k}\log\left(1+r_{ks}\right) \\ &+ \sum_{k\in\{1,2,4,5\}}\lambda_k^{(\mathbf{P})}\left(P_{k,max} - \sum_{l\in\mathcal{P}_k}\mathbf{1}^T\mathbf{P}_l\right) \\ &+\lambda_1^{(\mathbf{r})}\left(\sum_{n=1}^{N_F}c_{1n} - r_{12}\right) + \lambda_3^{(\mathbf{r})}\left(\sum_{n=1}^{N_F}c_{3n} - r_{11}\right) \\ &+\lambda_2^{(\mathbf{r})}\left(\sum_{n=1}^{N_F}c_{2n} - \left(r_{21}+r_{41}+r_{12}\right)\right) \\ &+\lambda_4^{(\mathbf{r})}\left(\sum_{n=1}^{N_F}c_{4n} - r_{41}\right) + \lambda_5^{(\mathbf{r})}\left(\sum_{n=1}^{N_F}c_{5n} - r_{22}\right) \\ &+\lambda_6^{(\mathbf{r})}\left(\sum_{n=1}^{N_F}c_{6n} - \left(r_{52}+r_{42}\right)\right) \\ &+\lambda_7^{(\mathbf{r})}\left(\sum_{n=1}^{N_F}c_{7n} - \left(r_{42}+r_{11}\right)\right) \\ &+\lambda_8^{(\mathbf{r})}\left(\sum_{n=1}^{N_F}c_{8n} - \left(r_{51}+r_{22}+r_{11}\right)\right) \\ &+\sum_{k,n,i}\lambda_{kni}^{(\mathbf{MUD})}\log\left(1+\frac{\sum_{\{j\in\mathcal{J}_i\}}\left|h_{jn}(t)\right|^2 P_{jn}}{\sigma^2}\right) \\ &-\lambda_{kni}^{(\mathbf{MUD})}\sum_{\{j\in\mathcal{J}_i\}}c_{jn}, \end{aligned}$$

where the sets of interfering links are: $\mathcal{I}_1 = \phi$ (i.e. empty set), $\mathcal{I}_2 = \{1,4\}$, $\mathcal{I}_3 = \{2,6\}$, $\mathcal{I}_4 = \{3\}$, $\mathcal{I}_5 = \{5,7\}$, and $\mathcal{I}_6 = \{8\}$; the vector $\mathbf{c} \triangleq \{c_{ln} : 1 \leq l \leq 8, 1 \leq n \leq 2\} \succeq \mathbf{0}$; and the vector $\boldsymbol{\lambda}$ is the collection of $\left\{\lambda_l^{(\mathbf{r})}\right\}$, $\left\{\lambda_k^{(\mathbf{P})}\right\}$ and $\left\{\lambda_{kni}^{(\mathbf{MUD})}\right\}$. Observed that interfering transmission links towards a common receiving node are coupled together via the link constraints to achieve a MUD capacity. Hence, we can partition the network via links in different MUD blocks. Specifically, the adaptive scaling matrix can be chosen with the following structure,

$$\mathbf{D}(t) = \text{blkdlg}\{\mathbf{D}_{\{\mathbf{MUD},1\}}(t), \mathbf{D}_{\{\mathbf{MUD},2\}}(t), \cdots, \mathbf{D}_{\{\mathbf{MUD},K\}}(t)\} \quad (30)$$

where $\{\mathbf{MUD}, k\}$ is the collection of variables coupled at the MUD node $k$. For example, $\{\mathbf{MUD}, 1\} = \{r_{11}, r_{22}, \lambda_1^{(\mathbf{P})}\}$, $\{\mathbf{MUD}, 2\} = \{r_{21}, r_{22}, c_{12}, c_{41}, p_1, p_4, \lambda_{2,1,i}^{(\mathbf{MUD})}, \lambda_1^{(\mathbf{r})}, \lambda_4^{(\mathbf{r})}, \lambda_2^{(\mathbf{P})}\}$ for the network realization in Fig. 1. $\mathbf{D}_{\{\mathbf{MUD},k\}}(t)$ is the adaptive scaling matrix associated with the collection of variables $\{\mathbf{MUD}, k\}$.

Using the solution in (29) and the block diagonal structure of $\mathbf{D}(t)$ in (30), the adaptive scaling matrix at MUD node $k$ is given by $\mathbf{D}^*_{\{\mathbf{MUD},k\}}(t+\overline{T}) = -\left(\nabla^2_{\{\mathbf{MUD},k\}}\mathcal{L}\left(\mathbf{r}(t), \mathbf{P}(t), \boldsymbol{\lambda}(t)\right)\right)^{-1}$ where the computation does not require global information but only local information from the neighboring nodes.

## V. Numerical Results and Discussions

In this section, we shall compare the PDSGA with three baseline schemes: (I) Bru-PDUA, i.e., primal-dual update algorithm with scaling matrices obtained by brute-force solving the optimization problem (27) with (26); (II) Dia-PDUA, i.e., primal-dual update algorithm with *diagonal* scaling matrices, whose diagonal entries are given by the diagonal entries of the corresponding Hessian matrix [23]; (III) Con-PDUA, i.e., the regular primal-dual update algorithm with constant stepsize $\xi = 0.005$ [23]. We also show the performance and constraint outage probability tradoff under different channel fading rates. Our simulations are based on the MCFC-MCPA example described in Section II-C and further analyzed in Section IV-B. Fig. 1 shows the network topology of the problem, where the network consists of $K = 6$ nodes and $L = 8$ directed links, and Fig. 2 shows the routing table. We consider one-hop data transmission. In addition, the average receiving SNR per-subband is 10 dB, the number of subband $N_F = 2$, and the number of data streams of a source node $R_k = 2$.

### A. Tracking Performance Comparison

Fig. 4 illustrates the normalized sum-utility versus time-slot index for the PDSGA and the three baseline schemes. As illustrated, the PDSGA using the proposed dynamic scaling matrix has a much better tracking capability than the baseline schemes Con-PDUA and Dia-PDUA, which are designed for quasi-static CSI. On the other hand, the PDSGA has similar performance as the brute-force solution Bru-PDUA, which shows that the approximation in (28) indeed works well. We also compare $\beta$ among the proposed PDSGA and the three baselines in Table I. The results also indicated that our algorithm has a better performance than Con-PDUA and Dia-PDUA.

### B. Region Stability Property

Fig. 5 shows the simulation results of region stability. The simulation results are consistent with the analytical results stated in *Theorem 1*, i.e., the probability that the algorithm trajectory at steady state being out of the limit region $\mathcal{Y}$ (see equation (22)) is proportional to the normalized update interval $\overline{T}/\overline{N}$. Moreover, as the scaling matrices in the PDSGA are adaptive to the time-varying CSI, the PDSGA performs better than the baseline schemes Con-PDUA and Dia-PDUA. On the other hand, the performance of the PDSGA and the brute-force solution Bru-PDUA are similar.

TABLE I
Comparison of Contraction Modulus $\beta$ in Different Algorithms over All State of CSI $\mathbf{h}^{(q)}$

| Algorithm | Average $\beta$ | Worst Case $\beta$ |
|---|---|---|
| Bru-PDUA | 0.3116 | 0.7862 |
| Con-PDUA | 0.9904 | 0.9998 |
| Dia-PDUA | 0.9401 | 0.9865 |
| PDSGA with proposed scaling matrix | 0.5876 | 0.8309 |

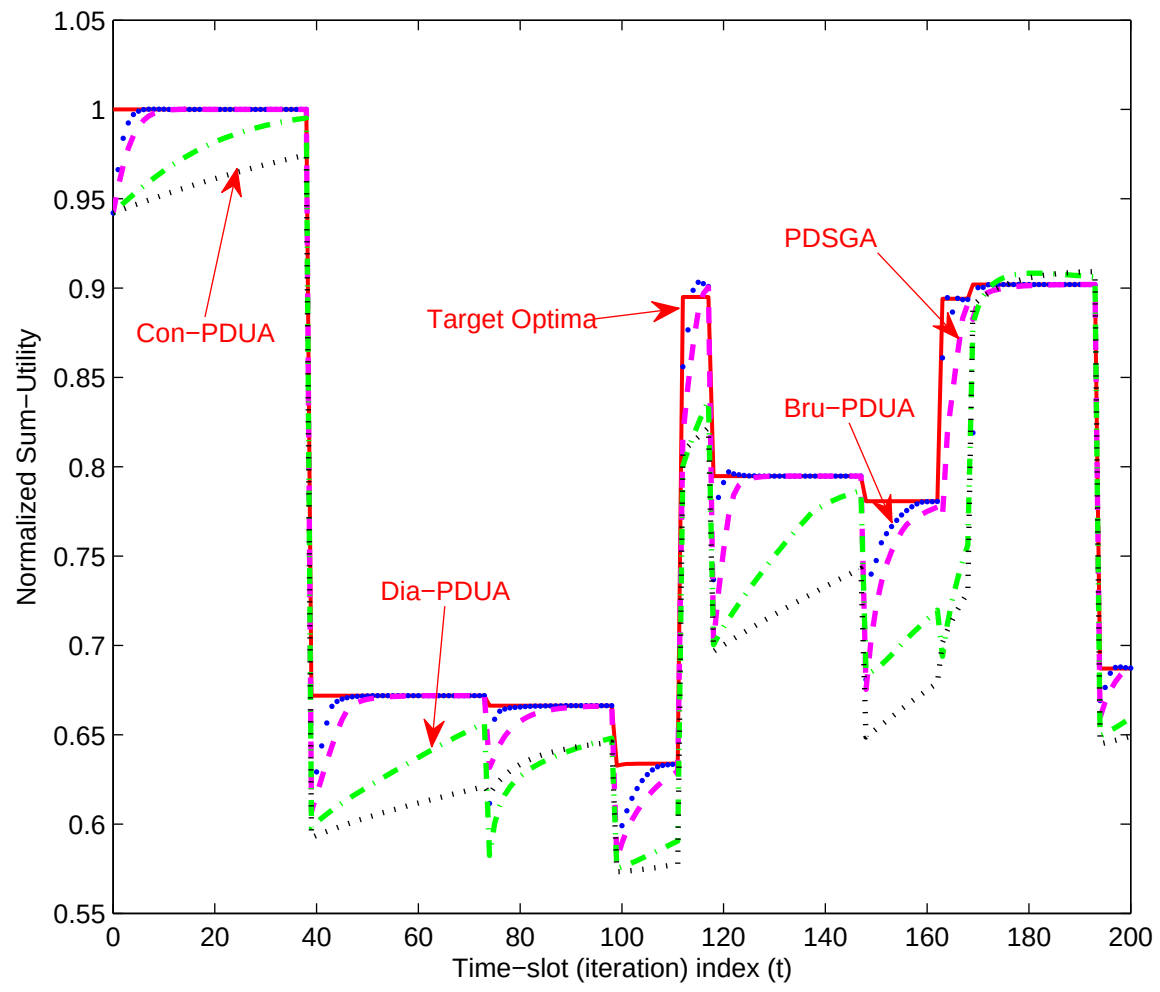

Fig. 4. Tracking performance comparison of the PDSGA and the baseline schemes: Bru-PDUA, Dia-PDUA and Con-PDUA, where Bru-PDUA denotes the primal-dual update algorithm with scaling matrices obtained by brute-force solving the optimization problem (27) with (26); Dia-PDUA denotes the primal-dual update algorithm with *diagonal* scaling matrices, whose diagonal entries are given by the diagonal entries of the corresponding Hessian matrix [23]; and Con-PDUA denotes the primal-dual update algorithm with constant stepsize $\xi = 0.005$ [23]. The sum-utility is normalized to the maximum sum-utility obtained across all the time-slots.

### C. Order of Growth of the Tracking Errors

Fig. 6 and Fig. 7 show the simulation results of the expected-absolute-error (EAE) and the mean-square-error (MSE), respectively. Both figures are consistent with the analytical results given in *Theorem 2*, i.e., the tracking errors, namely EAE and MSE, are proportional to the normalized update interval $\overline{T}/\overline{N}$. Moreover, as the scaling matrices in the PDSGA are adaptive to the time-varying CSI, the tracking errors associated with the PDSGA are much smaller than the baseline schemes Con-PDUA and Dia-PDUA. On the other hand, the performance of the PDSGA and the brute-force solution Bru-PDUA are quite similar.

### D. Constraint Outage Probability and Performance Tradeoff

Fig. 8 shows the tradeoff between constraint outage probability and performance under different channel fading rates.

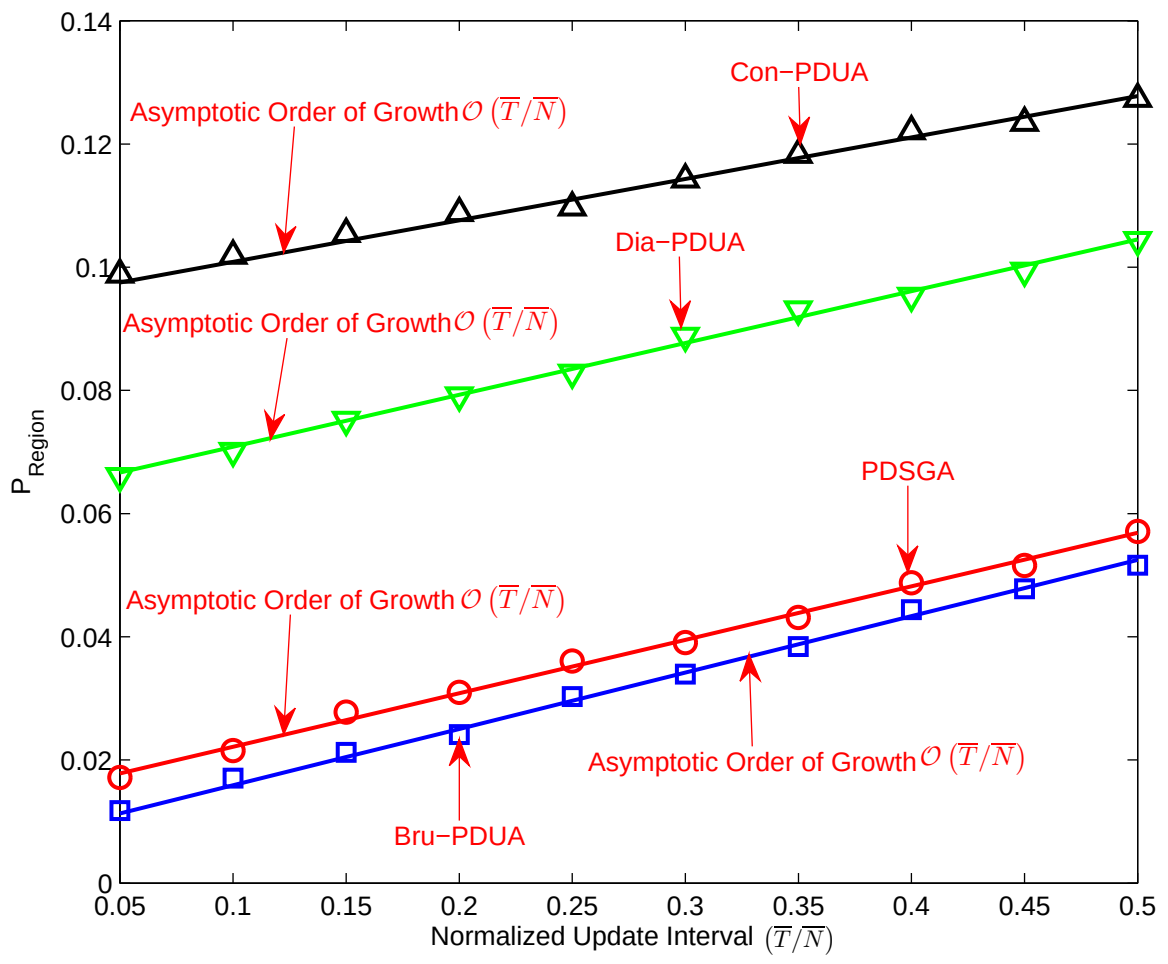

Fig. 5. Region stability property of the PDSGA and the baseline schemes: Bru-PDUA, Dia-PDUA and Con-PDUA, where Bru-PDUA denotes the primal-dual update algorithm with scaling matrices obtained by brute-force solving the optimization problem (27) with (26); Dia-PDUA denotes the primal-dual update algorithm with *diagonal* scaling matrices, whose diagonal entries are given by the diagonal entries of the corresponding Hessian matrix [23]; and Con-PDUA denotes the primal-dual update algorithm with constant stepsize $\xi = 0.005$ [23]. The asymptotic order of growth refers to the results stated in *Theorem 1* (see equation (21)).Here $\mathrm{P}_{\mathbf{Region}} = 1 - \mathrm{P}_{\mathcal{Y}}$.

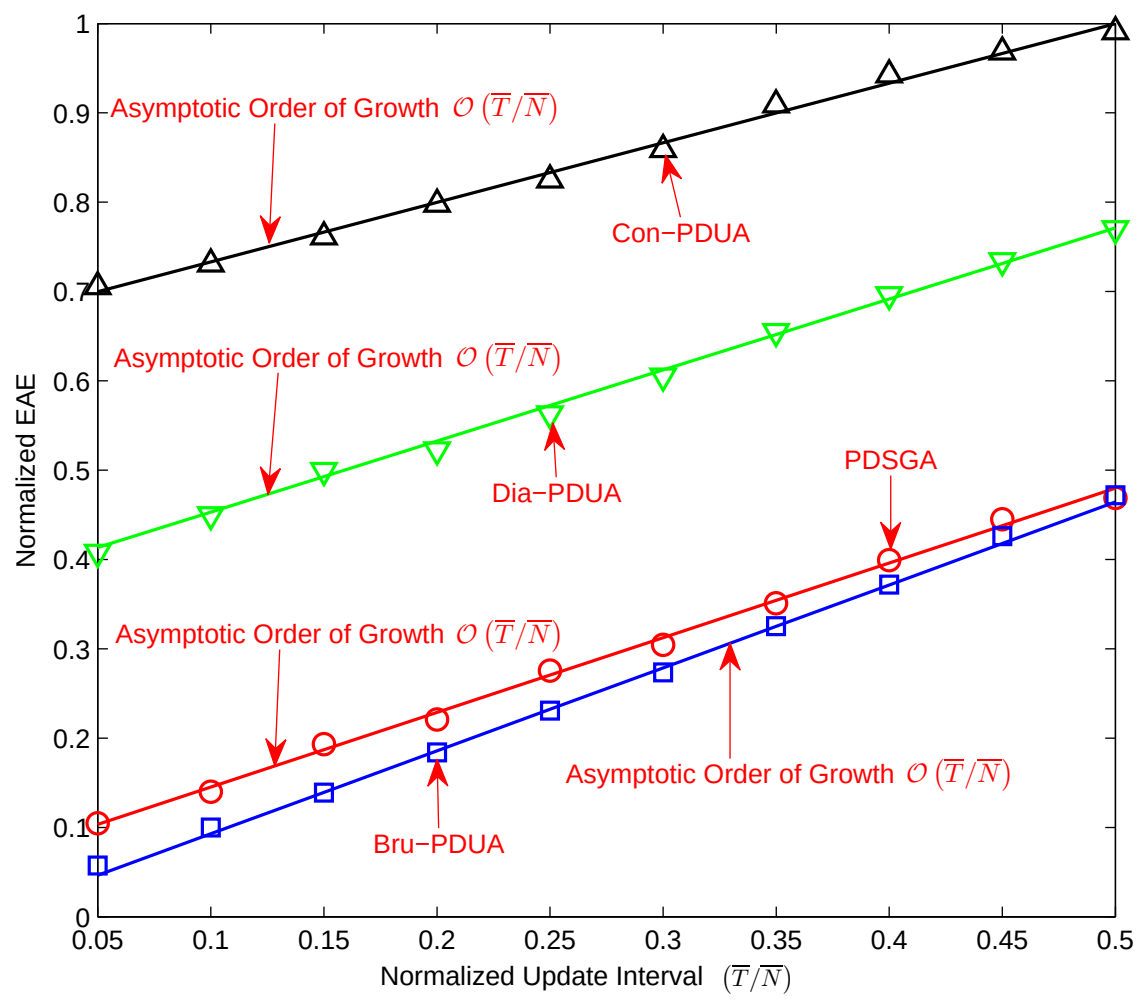

Fig. 6. Expected-absolute-error (EAE) versus Normalized Update Interval of the PDSGA and the baseline schemes: Bru-PDUA, Dia-PDUA and Con-PDUA [23]. The asymptotic order of growth refers to the results stated in *Theorem 2* (see equation (24)). The EAE is normalized to the maximum EAE of all the schemes.

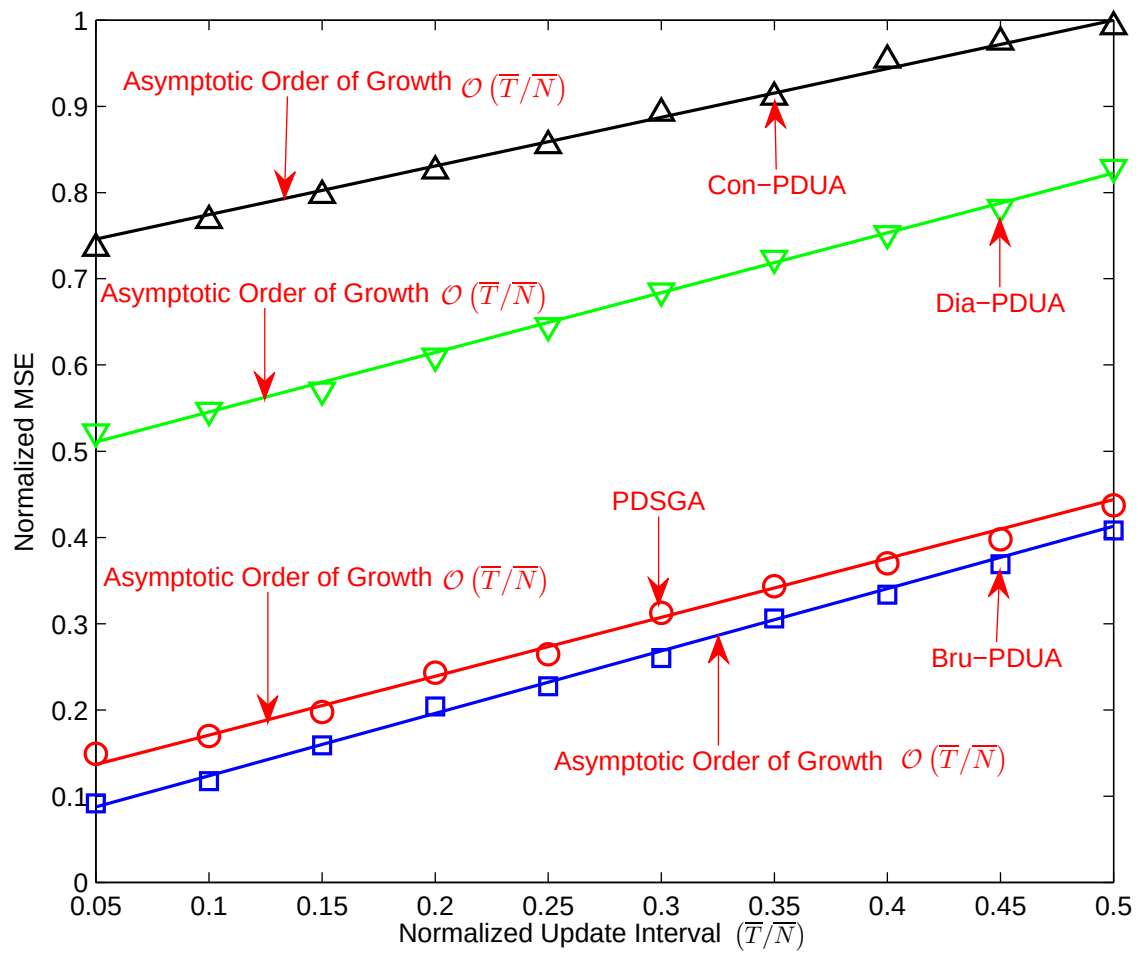

Fig. 7. Mean-square-error (MSE) versus Normalized Update Interval of the PDSGA and the baseline schemes: Bru-PDUA, Dia-PDUA and Con-PDUA [23]. The asymptotic order of growth refers to the results stated in *Theorem 2* (see equation (24)). The MSE is normalized to the maximum MSE of all the schemes.

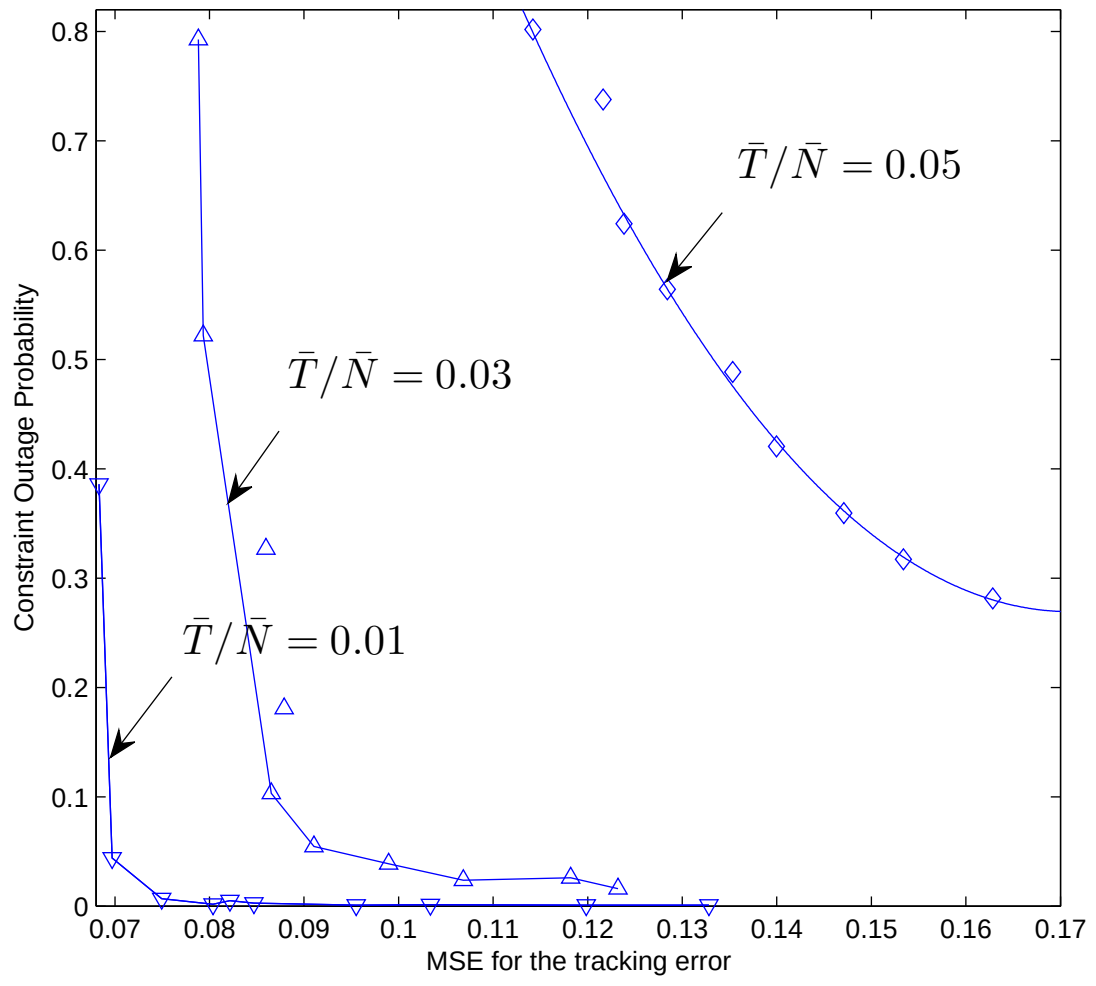

Fig. 8. Constraint Outage Probability versus Mean Square Error (MSE). The curves show a tradeoff between the Constraint Outage Probability and the performance in terms of the MSE under different channel fading rates which are in terms of $\bar{T}/\bar{N}$.

For slow varying CSI, (i.e. $\bar{T}/\bar{N}$ is small), it is easy to maintain a low constraint outage probability while achieve a good performance. However, for fast varying CSI, (i.e. $\bar{T}/\bar{N}$ is large), the performance has to be sacrificed to meet with some target constraint outage probability by introducing a large backoff margin according to equation (25).

## VI. Conclusions

In this paper, we have analyzed the convergence behavior of a primal-dual scaled gradient algorithm (PDSGA) that solves a network utility maximization (NUM) problem under time-varying fading channels. We have shown that the general PDSGA converges to a limit region rather than a single point under the FSMC model. Furthermore, the asymptotic tracking errors, namely the expected-absolute-error (EAE) and the mean-square-error (MSE), are both $\mathcal{O}\left(\overline{T}/\overline{N}\right)$. We have also analyzed the constraint outage probability due to time varying CSI and introduced a backoff margin to the constraint to reduce the outage probability. Based on these analysis, we have proposed a novel low complexity distributive algorithm to dynamically adjust the scaling matrix based on the time varying CSI in the PDSGA. Simulations have been carried out to verify the analytical results as well as to demonstrate the superior performance of the PDSGA over the baseline schemes.

## APPENDIX A
## PROOF OF *Lemma 1*

To prove *Lemma 1*, namely the contraction property of the PDSGA, we start with the left hand side (LHS) of equation (20) directly, i.e.,

$$\begin{aligned}&\left\|\mathbf{y}\left(t+\overline{T}\right)-\mathbf{y}_q^*\right\|_2\\&=\left\|\left[\mathbf{y}(t)+\mathbf{D}\left(\mathbf{h}^{(q)}\right)\nabla\mathcal{L}\left(\mathbf{y}(t);\mathbf{h}^{(q)}\right)\right]^+-\mathbf{y}_q^*\right\|_2\\&\overset{(a)}{\le}\left\|\left(\mathbf{y}(t)-\mathbf{y}_q^*\right)\right.\\&\qquad\left.+\mathbf{D}\left(\mathbf{h}^{(q)}\right)\left(\nabla\mathcal{L}\left(\mathbf{y}(t);\mathbf{h}^{(q)}\right)-\nabla\mathcal{L}\left(\mathbf{y}_q^*;\mathbf{h}^{(q)}\right)\right)\right\|_2,\end{aligned}$$

where $\overset{(a)}{\le}$ follows from the *non-expansive* property of the projection $[\cdot]^+$ operation [23]. Moreover, using mean-value theorem [23],

$$\begin{aligned}&\left\|\nabla\mathcal{L}\left(\mathbf{y}(t);\mathbf{h}^{(q)}\right)-\nabla\mathcal{L}\left(\mathbf{y}_q^*;\mathbf{h}^{(q)}\right)\right\|\\&\qquad=\left\|\nabla^2\mathcal{L}\left(\hat{\mathbf{y}};\mathbf{h}^{(q)}\right)\right\|\left\|\mathbf{y}(t)-\mathbf{y}_q^*\right\|,\end{aligned}$$

where $\nabla^2\mathcal{L}\left(\hat{\mathbf{y}}(t);\mathbf{h}^{(q)}\right)\in\mathbb{R}_+^{(M+N_c)\times(M+N_c)}$ denotes the *second-order* derivative of the Lagrangian $\mathcal{L}(\mathbf{y}(t);\,\mathbf{h}^{(q)})$ w.r.t. $\mathbf{y}$, evaluated at $\hat{\mathbf{y}}(t)$; and $\hat{\mathbf{y}}(t)$ is given by $\hat{\mathbf{y}}(t)=\mathbf{y}(t)+(1-\varpi)\mathbf{y}_q^*$, with $\varpi\in[0,1]$. Hence we obtain

$$\begin{aligned}&\left\|\mathbf{y}\left(t+\overline{T}\right)-\mathbf{y}_q^*\right\|_2\\&=\left\|\left(\mathbf{I}+\mathbf{D}\left(\mathbf{h}^{(q)}\right)\nabla^2\mathcal{L}\left(\hat{\mathbf{y}};\mathbf{h}^{(q)}\right)\right)\left(\mathbf{y}(t)-\mathbf{y}_q^*\right)\right\|_2\\&\overset{(b)}{\le}\left\|\mathbf{I}+\mathbf{D}\left(\mathbf{h}^{(q)}\right)\nabla^2\mathcal{L}\left(\hat{\mathbf{y}};\mathbf{h}^{(q)}\right)\right\|_2\left\|\mathbf{y}(t)-\mathbf{y}_q^*\right\|_2\\&\overset{(c)}{\le}\beta_q\left\|\mathbf{y}(t)-\mathbf{y}_q^*\right\|_2,\end{aligned}$$

where $\overset{(b)}{\le}$ follows from the property of the standard matrix-2 norm; and $\overset{(c)}{\le}$ follows from the contraction condition given in equation (19).

## APPENDIX B
## PROOF OF *Theorem 1*

Consider a time interval $[0,\ T]$ with $n$ switchings, i.e., there are $n$ stages in the interval $[0,\ T]$. Let $\{q_1,q_2,\cdots,q_n\}$, $\{T_1,T_2,\cdots,T_n\}$ and $\{\beta_1,\beta_2,\cdots,\beta_n\}$ denote the channel states, (random) sojourn times and contraction modulus of the $n$ stages, respectively, as shown in Fig. 9.

Under the conditions that $\beta\left(\mathbf{D}(m)\right)<1$, the iteration (17) is contraction mapping in the $m^{th}$ stage ($1\le m\le n$) [23]. As a result, the distances $\{\left\|\mathbf{e}(m)\right\|_2,1\le m\le n\}$ between the iterate $\mathbf{y}(m)$ and the target optimal solution at the end of each stage can be bounded by

$$\begin{aligned}&\left\|\mathbf{e}(1)\right\|_2\le\left\|\mathbf{y}(0)\right\|\beta_1^{N_1};\\&\left\|\mathbf{e}(m)\right\|_2\le\left\|\mathbf{e}(m-1)\right\|_2\beta_m^{N_m}+\delta_{m-1,m}\end{aligned}\qquad(31)$$

where $\delta_{m-1,m}$ denotes the distance between the target optimal solution of the $(m-1)^{th}$ stage and $m^{th}$ stage, i.e., the jump of the equilibrium point of the switched system; and $N_m$ denotes the number of updates during the $m^{th}$ stage. Iterating equation (31) from $m=1$ to $m=n$, we can get

$$\begin{aligned}\|e(n)\|_2&\le\left\{\left[\|\mathbf{y}(0)\|\beta_1^{N_1}+\delta_{1,2}\right]\beta_2^{N_2}+\delta_{2,3}\right\}\beta_3^{N_3}+\cdots\\&=\|\mathbf{y}(0)\|\prod_{m=1}^{n}\beta_m^{N_m}+\sum_{l=1}^{n}\prod_{m=l}^{n}\delta_{m-1,m}\beta_m^{N_m}\\&\le\|\mathbf{y}(0)\|\beta^{\sum_{m=1}^{n}N_m}+\delta\sum_{l=2}^{n}\prod_{m=l}^{n}\beta^{N_m}\end{aligned}\qquad(32)$$

where $\delta=\max_m\{\delta_{m-1,m}\}$ is the maximum distance between any two equilibrium points.

Let $\mathbf{q}=[q_1\ \ q_2\ \ \cdots\ \ q_n]$ and $\Gamma=[T_1\ \ T_2\ \ \cdots\ \ T_n]$, then take expectation of both sides of equation (32) w.r.t. $\{\mathbf{q},\Gamma\}$, we can get

$$\begin{aligned}\mathbb{E}_{\{\mathbf{q},\Gamma\}}\left\{\|\mathbf{e}(n)\|_2\right\}&\le\mathbb{E}_{\{\mathbf{q},\Gamma\}}\left\{\mathbf{y}(0)\beta^{\sum_{m=1}^{n}N_m}\right\}\\&+\mathbb{E}_{\{\mathbf{q},\Gamma\}}\left\{\delta\sum_{l=2}^{n}\prod_{m=l}^{n}\beta^{N_m}\right\}.\end{aligned}\qquad(33)$$

Under the assumption that $[\mathbf{T}]_{qq}=\nu^{K^2},\forall\ q\in\mathcal{Q}$ (see equation (3)), then we know that $T_1,T_2,\cdots,T_n$ are identically distributed with probability mass function (PMF) given by

$$\Pr\{N_m=l\}=\nu^{L^2(l-1)}\left(1-\nu^{L^2}\right),\ \forall\ l=1,2,\cdots.$$

Then, we have

$$\begin{aligned}\mathbb{E}_{\{\mathbf{q},\Gamma\}}\left\{\delta\sum_{l=2}^{n}\prod_{m=l}^{n}\beta^{N_m}\right\}&=\mathbb{E}_{\{\mathbf{q}\}}\left\{\mathbb{E}_{\{\Gamma\}}\left\{\delta\sum_{l=2}^{n}\prod_{m=l}^{n}\beta^{N_m}\middle|\mathbf{q}\right\}\right\}\\&=\frac{\delta\rho\left(1-\rho^{n-1}\right)}{1-\rho},\end{aligned}\qquad(34)$$

where $\rho=\mathbb{E}\left\{\beta^{N_m}\right\}=\frac{\overline{T}\beta\left(1-\nu^{L^2}\right)}{1-\beta\nu^{K^2}}=\frac{\overline{T}\beta}{\beta+(1-\beta)\overline{N}}$. Therefore, let $n\to+\infty$, we get

$$\mathbb{E}_{\{\mathbf{q},\Gamma\}}\left\{\|\mathbf{e}(\infty)\|_2\right\}\le\frac{\delta\alpha}{1-\alpha}=\frac{\overline{T}\delta\beta}{(1-\beta)\overline{N}}.\qquad(35)$$

Then, by virtue of the Markov inequality we can get $\Pr\left\{\|\mathbf{e}(\infty)\|_2>\delta\right\}\le\frac{\beta\overline{T}}{(1-\beta)\overline{N}}$, which means $\lim_{n\to+\infty}\Pr\left\{\mathbf{y}(t)\in\mathcal{Y}\right\}\ge1-\min\left\{1,\frac{\beta\overline{T}}{(1-\beta)\overline{N}}\right\}$.

## APPENDIX C
## PROOF OF *Theorem 2*

As in *Appendix B*, we consider a time interval $[0,\ T]$ with $n$ switchings, i.e., there are $n$ stages in the interval $[0,\ T]$. Let $\{q_1,q_2,\cdots,q_n\}$, $\{T_1,T_2,\cdots,T_n\}$ and $\{\beta_1,\beta_2,\cdots,\beta_n\}$ denote the channel states, (random) sojourn times and contraction modulus of the $n$ stages, respectively, as shown in Fig. 9. From equation (35) in *Appendix B* we know that, at steady state (i.e., when $t\to+\infty$), EAE $(\mathbf{y}(t))\le\frac{\delta\beta\overline{T}}{(1-\beta)\overline{N}}$.

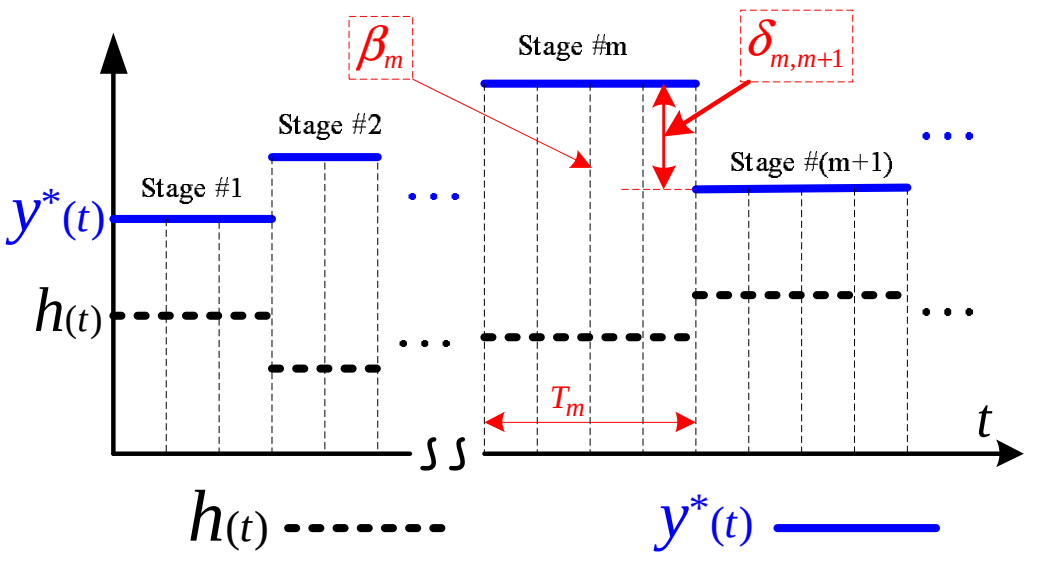

Fig. 9. An illustration of the concept *stage* of channel states and the associated *target optimal solution* in one dimensional space [11]. The length $T_m$ of the $m^{th}$ stage is a random variable, which can be one or several time-slots ; and the number of updates during the $m^{th}$ stage is denoted as $n_m$; the distance between the target optimal solution of the $m^{th}$ stage $\mathbf{y}_m^*$ and the target optimal solution of the $(m+1)^{th}$ stage $\mathbf{y}_{(m+1)}^*$ is denoted as $\delta_{m,m+1}$; and the contraction modulus of in the $m^{th}$ stage is denoted as $\beta_m$, $\forall m \geq 1$ [11].

Let $\eta = \mathbb{E}\left\{\beta^{2n_m}\right\} = \frac{\beta^2\left(1-\nu^{L^2}\right)}{1-\beta^2\nu^{L^2}} = \frac{\beta^2\overline{T}}{\beta^2+(1-\beta^2)\overline{N}}$, we then have

$$\begin{aligned}
&\mathbb{E}_{\{\mathbf{q},\Gamma\}}\left\{\left(\delta\sum_{l=2}^{n}\prod_{m=l}^{n}\beta^{n_m}\right)^2\right\} \\
&= \mathbb{E}_{\{\mathbf{q}\}}\left\{\mathbb{E}_{\{\Gamma\}}\left\{\left(\delta\sum_{l=2}^{n}\prod_{m=l}^{n}\beta^{n_m}\right)^2\Bigg|\mathbf{q}\right\}\right\} \\
&= \frac{2\eta}{1-\eta}\left[\frac{1-\rho^{n-1}}{1-\rho} - \frac{\eta^n-\eta\rho^{n-1}}{\eta-\rho}\right] - \frac{\eta(1-\eta^{n-1})}{1-\eta}.
\end{aligned}$$

Combining the above equation with equation (33), and letting $n \to +\infty$, we can get

$$\text{MSE}\left(\mathbf{y}(t)\right) \leq \frac{\delta^2\beta^2\left(2\beta+\overline{T}(1-\beta)\overline{N}\right)}{(1-\beta^2)(1-\beta)\overline{N}^2}.$$

## Appendix D
## Proof of *Lemma 2*

To introduce a backoff margin $K$ to reduce the constraint outage probability, instead of keeping the constraint $g(\mathbf{x}) \leq 0$, we form the constraint as

$$g(\mathbf{x}) + K \leq 0.$$

Hence at the equilibrium point $\mathbf{x}^*(t)$, it is satisfied that $g(\mathbf{x}^*(t)) \leq -K$ and we have the relationship

$$g(\mathbf{x}(t)) > \epsilon \Rightarrow g(\mathbf{x}(t)) - g(\mathbf{x}^*(t)) > \epsilon + K,$$

which implies that

$$\Pr\{g(\mathbf{x}(t)) > \epsilon\} \leq \Pr\{g(\mathbf{x}(t)) - g(\mathbf{x}^*(t)) > \epsilon + K\}. \quad (36)$$

From condition (ii), since $\|g(\mathbf{x}(t)) - g(\mathbf{x}^*(t))\| \leq c\|\mathbf{x}(t) - \mathbf{x}^*(t)\|$, we obtain

$$\begin{aligned}
&\Pr\{g(\mathbf{x}(t)) - g(\mathbf{x}^*(t)) > \epsilon + K\} \\
&\leq \Pr\{\|g(\mathbf{x}(t)) - g(\mathbf{x}^*(t))\| > \epsilon + K\} \\
&\leq \Pr\{\|\mathbf{x}(t) - \mathbf{x}^*(t)\| > \frac{\epsilon + K}{c}\} \qquad (37)
\end{aligned}$$

Taking the intermediate result from the proof of *Theorem 1* (see equation (35)), we finally get

$$\lim_{t\to\infty}\Pr\{\|\mathbf{x}(t) - \mathbf{x}^*(t)\| > \frac{\epsilon+K}{c}\} \leq \frac{c\delta\beta\bar{T}}{(\epsilon+K)(1-\beta)\bar{N}} \quad (38)$$

Combining (36)-(38) we complete the proof.

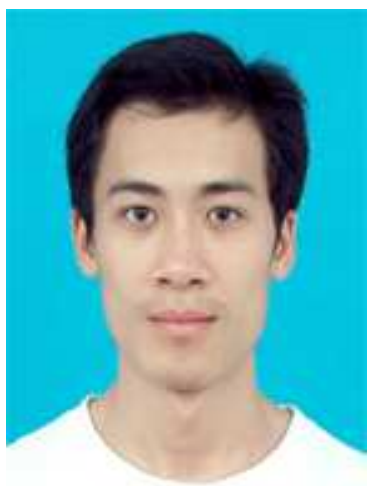

**Junting Chen** received the B.Sc. (1st Hons.) in electronic science and technology from Nanjing University, Nanjing, China, in 2009.

He is currently working toward the Ph.D. degree at the Department of Electronic and Computer Engineering, The Hong Kong University of Science and Technology (HKUST). His research interests include resource allocation and convex optimization in time-varying wireless channels.

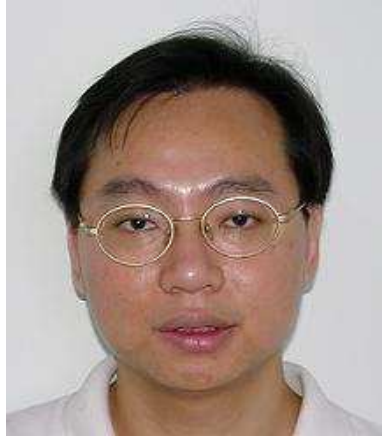

**Vincent K. N. Lau** received the B.Eng. (Distinction 1st Hons.) from the University of Hong Kong in 1992 and the Ph.D. degree from Cambridge University, Cambridge, U.K., in 1997.

He was with HK Telecom (PCCW) as a System Engineer from 1992 to 1995, and with Bell Labs - Lucent Technologies as a member of Technical Staff during 1997-2003. He then joined the Department of ECE, HKUST, and is currently a Professor. His current research interests include the robust and delay-sensitive cross-layer scheduling of MIMO/OFDM wireless systems, cooperative and cognitive communications, dynamic spectrum access, as well as stochastic approximation and Markov decision process.

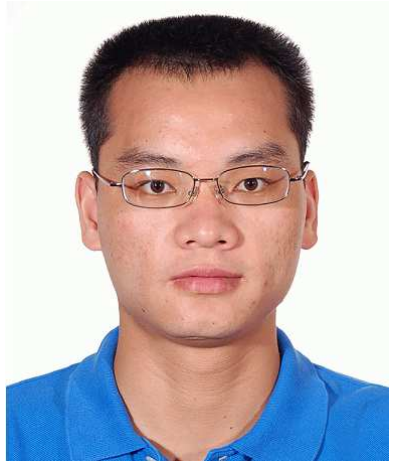

**Yong Cheng** received the B.Eng. (1st hons.) from Zhejiang University (2002-2006), Hangzhou, China, and the master of philosophy degree in 2010 from the department of Electrical and Computer Engineering, The Hong Kong University of Science and Technology (HKUST).

His recent research interests include resource control and optimization in wireless networks and cooperative MIMO systems.